\RequirePackage{fix-cm}
\documentclass[smallextended]{svjour3}       
\smartqed  
\usepackage[hyphens]{url}
\usepackage{natbib}
\usepackage{amsmath,amssymb,amsfonts}
\usepackage{algorithmic}
\usepackage{graphicx, subfigure}
\usepackage{textcomp}
\usepackage{xcolor}
\usepackage[skins,breakable]{tcolorbox}
\usepackage{booktabs}
\usepackage[htt]{hyphenat}
\usepackage[hidelinks]{hyperref}
\usepackage{float}
\usepackage{caption}
\usepackage[misc]{ifsym}
\usepackage{todonotes}
\usepackage{listings}
\usepackage[normalem]{ulem}
\usepackage{mathptmx}
\usepackage{pifont}

\usepackage{makecell}
\usepackage{soul}
\usepackage{listings}
\usepackage{multirow}
\usepackage{pifont}
\usepackage{xspace}
\usepackage[flushleft]{threeparttable}
\usepackage[T1]{fontenc}
\usepackage{lscape}
\usepackage{afterpage}
\usepackage{wrapfig}
\definecolor{backcolour}{rgb}{0.95,0.95,0.92}

\def\BibTeX{{\rm B\kern-.05em{\sc i\kern-.025em b}\kern-.08em
    T\kern-.1667em\lower.7ex\hbox{E}\kern-.125emX}}
    
\graphicspath{{figs/}}

\newcommand{\fix}[1]{\textcolor{black}{#1}}

\newcommand{\minor}[1]{\textcolor{black}{#1}}

\newcommand{\RqOne}{\textbf{(RQ1)} \ul{\textit{How prevalent are clones of self-admitted technical debt in build systems?}}\xspace}

\newcommand{\RqTwo}{\textbf{(RQ2)} \ul{\textit{How prevalent is cloning of the statements that surround clones of self-admitted technical debt in build systems?}} \xspace} 

\newcommand{\RqThree}{\textbf{(RQ3)} \ul{\textit{Which developers are authoring clones of self-admitted technical debt in build systems?}} \xspace} 

\newcommand{\RqFour}{\textbf{(RQ4)} \ul{\textit{What are the characteristics of clones of self-admitted technical debt in build systems?}} \xspace}

\definecolor{Large}{HTML}{696969}
\definecolor{Negligible}{HTML}{D3D3D3}
\definecolor{Medium}{HTML}{808080}
\begin{document}

\title{Quantifying and Characterizing Clones of Self-Admitted Technical Debt in Build Systems}


\author{Tao Xiao~\Letter \and Zhili Zeng \and Dong Wang~\Letter  \and Hideaki Hata \and Shane McIntosh \and Kenichi Matsumoto
}


\institute{
    \Letter~Corresponding author - Tao Xiao
    \at Nara Institute of Science and Technology, Japan\\
    \email{tao.xiao.ts2@is.naist.jp}
    \and
    Zhili Zeng, Shane McIntosh
    \at University of Waterloo, Canada\\
    \email{\{z75zeng,shane.mcintosh\}@uwaterloo.ca}
    \and
    \Letter~Corresponding author - Dong Wang
    \at College of Intelligence and Computing, Tianjin University, China\\
    \email{d.wang@ait.kyushu-u.ac.jp}
    \and 
    Hideaki Hata
    \at Shinshu University, Japan\\
    \email{hata@shinshu-u.ac.jp}
    \and 
    Kenichi Matsumoto
    \at Nara Institute of Science and Technology, Japan\\
    \email{matumoto@is.naist.jp}
}

\date{Author pre-print copy. The final publication is available at Springer via:\\
\url{https://link.springer.com/article/10.1007/s10664-024-10449-5}}

\maketitle

\begin{abstract}
Self-Admitted Technical Debt (SATD) annotates development decisions that intentionally exchange long-term software artifact quality for short-term goals. 
Recent work explores the existence of SATD clones (duplicate or near duplicate SATD comments) in source code. Cloning of SATD in build systems (e.g., CMake and Maven) may propagate suboptimal design choices, threatening qualities of the build system that stakeholders rely upon (e.g., maintainability, reliability, repeatability).
Hence, we conduct a large-scale study on 50,608 SATD comments extracted from Autotools, CMake, Maven, and Ant build systems to investigate the prevalence of SATD clones and to characterize their incidences.  
We observe that: (i) prior work suggests that 41--65\% of SATD comments in source code are clones, but in our studied build system context, the rates range from 62\% to 95\%, suggesting that SATD clones are a more prevalent phenomenon in build systems than in source code;
(ii) statements surrounding SATD clones are highly similar, with 76\% of occurrences having similarity scores greater than 0.8;
(iii) a quarter of SATD clones are introduced by the author of the original SATD statements; 
and (iv) among the most commonly cloned SATD comments, external factors (e.g., platform and tool configuration) are the most frequent locations, limitations in tools and libraries are the most frequent causes, and developers often copy SATD comments that describe issues to be fixed later. 
Our work presents the first step toward systematically understanding
SATD clones in build systems and opens up avenues for future work, such as distinguishing different SATD clone behavior, as well as designing an automated recommendation system for repaying SATD effectively based on resolved clones.
\end{abstract}

\section{Introduction}
\label{intro}
Technical Debt (TD) is a metaphor that is used to describe the impact that suboptimal design choices have on a software project~\citep{cunningham1992wycash}. When implementing an optimal solution that conflicts with other project goals (e.g., time-to-release) or constraints (e.g., development cost), it is often prudent to select a suboptimal solution. In the TD metaphor, this is akin to taking out a loan. The suboptimal solution will generate costs for the development team, which equates to ``interest payments'' in the TD metaphor. Intentional TD is often denoted using code comments to document the reasoning behind design choices. The literature refers to such examples as Self-Admitted Technical Debt (SATD)~\citep{potdar2014exploratory}.

Researchers have explored how SATD tends to accrue in software development artifacts, such as source code~\citep{potdar2014exploratory, vidoni2021self} and the software development process~\citep{xavier2020beyond, kashiwa2022empirical}.
In terms of source code, \citet{potdar2014exploratory} observed that SATD exists in 31\% of
source code files from four Java systems. 
\citet{vidoni2021self} found that 3\% of comments are SATD in more than 500 R packages.
Concerning the software development process, \citet{xavier2020beyond} discovered that only 29\% of the studied SATD in issue reports could be tracked to source code comments. 
Moreover, \citet{kashiwa2022empirical} demonstrated that at least 28\% of SATD is introduced during code reviews, and 20\% of SATD is introduced due to the suggestions of reviewers. 
Apart from source code and the software development process, our prior work~\citep{9551792} showed that comments in build systems (which orchestrate order- and configuration-dependent tool invocations to produce a repeatable and incremental process) also accrue SATD.

When developers encounter a familiar problem, it is not uncommon for them to copy and paste solutions, creating so-called software clones~\citep{roy2007survey}. Recent work~\citep{10.1145/3524610.3528387, 9551792} has shown that this practice of software cloning also applies to SATD.
For example, \citet{10.1145/3524610.3528387} pointed out that more than 40\% of SATD comments are duplicated in source code files, observing that the major root cause behind these SATD clones (duplicate or near-duplicate) is due to the need of cloning the code under the original SATD comment.
In the context of build systems, 
our prior work~\citep{9551792} revealed that the same SATD workaround could be applied across different projects. 
Moreover, \citet{10.1145/2591062.2591181} found that cloning is an even more pervasive practice in build systems than in source code.

Based on the prior work, we conjecture that SATD clones would also broadly exist in build systems.
Build systems play an important role at the heart of modern software development (e.g., test suite automation, external and internal dependency management)~\citep{kumfert2002software, mcintosh2011empirical}.
These clones may propagate technical debt throughout and across build systems, potentially hampering qualities of the build systems that stakeholders rely upon, such as maintainability, since repayments of the debt will need to be made consistently across clone copies.
For example, the cloned SATD comments from different specifications would record the same latent bugs or workarounds. 
If only one SATD instance is repaid, the other copies will become inconsistent and may still suffer from the symptoms of the underlying issue, accruing additional ``interest\fix{''}.
Therefore, it is important to not only effectively track and manage SATD, but also the clones of SATD in dependent artifacts such as in build systems.

To explore the extent to which cloning of SATD impacts build systems, we conduct an empirical study on 50,608 SATD comments from 3,427 GitHub projects, spanning four popular build systems (Autotools, CMake, Maven, and Ant). 
Specifically, we investigate the prevalence of SATD clones, explore the similarity of statements (a.k.a., lines of codes) that surround cloned SATD comments, analyze the authorship of SATD clones, and characterize the most commonly cloned SATD comments in build systems. 
We structure the paper by addressing the following four research questions:\\
\RqOne \\
\RqTwo \\
\RqThree \\
\RqFour \\

\fix{
Our research has yielded several significant findings related to SATD clones in build systems. First, we discovered that SATD comments are frequently cloned, with 75\% of clones found in external repositories. Notably, SATD comment clones in Autotools emerge as the most widespread. Second, while the statements surrounding SATD clones demonstrate high similarity, they score lower in similarity when compared to those surrounding non-SATD clones. Third, about one-fourth of SATD clones can be traced back to the author of the original SATD statements. This contrasts with the observation that non-SATD clones have more diverse authorship, yet a single author tends to clone SATD more frequently than non-SATD. Last, our analysis of 200 prevalent cloned SATD comments identified three distinct locations, eight reasons, and six underlying purposes. A recurrent theme is the interplay with external factors, often attributed to constraints in tools and libraries. It appears that developers tend to replicate these SATD comments predominantly to document challenges intended for future resolution.}

To summarize, the contributions of this work are three-fold:
(I) we systematically investigate the cloning phenomenon of SATD comments by taking into account diverse and popular build systems;
(II) we construct a taxonomy of characteristics of cloned SATD comments (i.e., locations, reasons, and purposes) in four build systems through a manual inspection to establish a broader theory of
SATD in build systems in general; and (III)
our findings, which highlight the prevalence of cloning in build systems, lay the foundation for an automatic SATD repayment system, from the actionable suggestions provided for the researchers and practitioners.



\section{Motivating example}
\label{mot}


\begin{figure}[t]
  \centering
  \includegraphics[width=.8\linewidth]{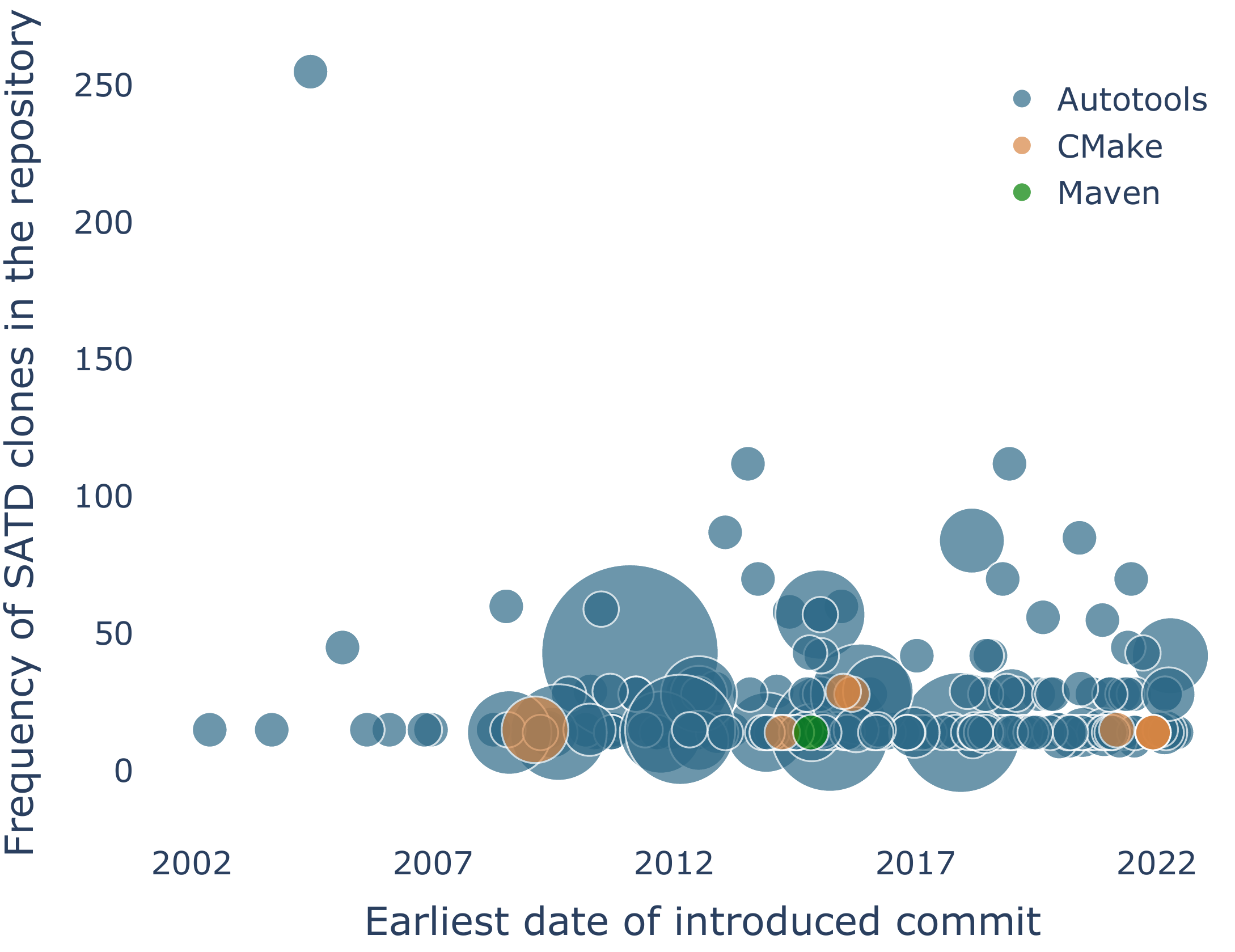}
      \caption{The propagation of a SATD comment ``\# FIXME: insert proper C++ library support''.}
  \label{fig:mot}
\end{figure}

We now illustrate an example to motivate our study.
Example~\ref{example1} is the most pervasively cloned SATD comment (5,413 times in 237 GitHub repositories) in our studied dataset:

\begin{lstlisting}[linewidth=\columnwidth,breaklines=true, showstringspaces=false, columns=fullflexible,basicstyle=\footnotesize, caption={The most pervasively cloned SATD comment.}, label={example1}]
AC_MSG_CHECKING([whether the $compiler linker ($LD) supports shared libraries])
    _LT_TAGVAR(ld_shlibs, $1)=yes
    case $host_os in
      aix3*)
        # FIXME: insert proper C++ library support
        _LT_TAGVAR(ld_shlibs, $1)=no
\end{lstlisting}
Figure~\ref{fig:mot} presents the propagation of this SATD comment via a bubble plot, where each bubble represents a GitHub repository and the size of the bubble denotes the popularity of a repository (i.e., stars).
We highlight two observations from this motivating example:

\noindent 
\textit{\ul{(I) SATD comment is cloned across multiple build tools.}}
As shown in the figure, this SATD comment (``\# FIXME: insert proper C++ library support'') is not only repeatedly cloned in a single build tool (e.g., Autotools), but also be cloned in repositories that contain other preliminary build tools (e.g., Maven and CMake).

\noindent
\textit{\ul{(II) SATD clones have long-life propagation.}}
In view of the timeline, Figure~\ref{fig:mot} shows that this temporary hack has been cloned from 2002 to 2022, accounting for more than ten years. 
This hack was first introduced on May 15, 2002, in \textit{open-watcom/open-watcom-v2} repository.
Specifically, we observe that since 2010, this SATD was rapidly propagated to many other repositories including those popular ones. 
For example, it occurs in one famous repository having 22,671 stars: \textit{emscripten-core/emscripten} which compiles C and C++ to \texttt{WebAssembly} using \texttt{LLVM} and \texttt{Binaryen}.

\section{Study Design}
\label{data}
In this section, we first propose the research questions with their motivations. We then introduce the studied projects and present the descriptions of extracting
SATD comments, identifying SATD clones, and preparing non-SATD clones.

\subsection{Research Questions}
To construct our paper, we formulate the following four research questions:

\noindent
\RqOne \\
\noindent
\citet{10.1145/3524610.3528387} found that a considerable proportion (41--65\%) of SATD being detected in Java files is cloned. This calls into question the true prevalence of SATD in Java files, and presents new challenges in how it can be tracked and managed. In the build system context, it is unclear the extent to which cloning is present in the SATD that is being detected. Therefore, we set out to first study the prevalence of SATD clones in build systems to understand whether similar tracking and management concerns are impacting build maintainers.

\noindent    
\RqTwo \\
\noindent
We assume that SATD clones are likely to occur due to cloned statements surrounding them. To evaluate this assumption, we set out to study cloning rates in the statements that surround SATD comments in build systems.
    
\noindent   
\RqThree \\
\noindent
\fix{Understanding the authorship of cloned SATD comments is crucial as it can pave the way for the development of automated awareness mechanisms in the future. Therefore, we sought to ascertain if the cloned SATD comments originate from identical or disparate authors, underlining the potential need for future tools that cater to SATD comments by analyzing the variances in SATD clones linked to different authors.}
    
\noindent
\RqFour \\
\noindent
Although our prior work~\citep{9551792} characterized SATD comments in the single Maven build system, it is still unclear what characteristics of pervasively cloned SATD comments \fix{are shared} in Autotools, CMake, Maven, and Ant build systems. 
    Answering this RQ could shed light on the future proposal of automatic SATD classifiers to manage SATD clones and eventually repay them.

\subsection{Studied Projects}
\label{sec:proj}
In order to draw conclusions about the nature of SATD clones in build systems, we construct a large-scale and diverse dataset. 
\minor{We initially adopt 735,669 projects from the dataset provided by \citet{dabic2021sampling}, which describes 25 characteristics of these projects.}
Referring to the prior study on build system clones~\citep{10.1145/2591062.2591181}, we then select the commonly studied build systems written in Java (Maven, Ant, and Ivy) and C/C++/Objective C projects (GNU Autotools and CMake) as our target.
In addition, we embrace Ivy in this study since it is an extension of Ant for managing external libraries.
Afterward, we are able to obtain 258,216 projects from the following four programming languages: 60,583 C projects, 77,193 C++ projects, 29,254 Objective C projects, and 91,186 Java projects.

\begin{figure}[t]
  \centering
  \includegraphics[width=.8\linewidth]{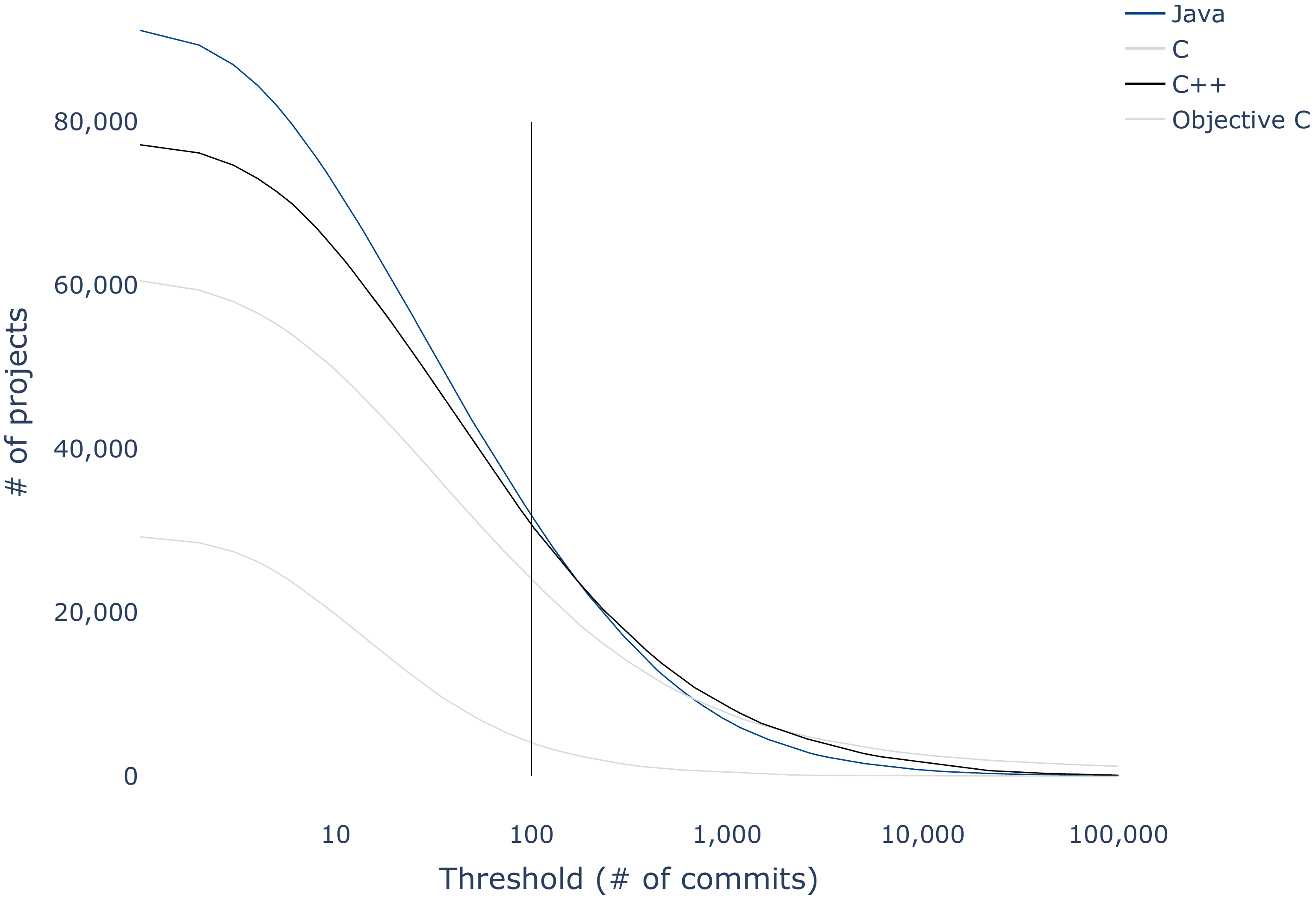}
  \caption{Threshold plot for the number of commits.}
  \label{fig:commits}
\end{figure}
To ensure that the empirical observations are significant, we employ four criteria for filtering out projects:
\begin{itemize}
    \item \textbf{C1}: The first criterion is to filter out those inactive projects based on the number of commits. \fix{In line with prior work~\citep{gallaba2018use,xiao202318}, we rely on threshold plots, depicted in Figure~\ref{fig:commits}, as a diagnostic tool. This assists us in selecting a threshold approximating the point of diminishing returns, leading us to set a threshold at 100 commits.}  In other words, we filter out those projects whose commits are less than 100.
    \item \textbf{C2}: During this filtering, we filter out those toy and recently dormant projects based on the following conditions \fix{from previous studies~\citep{kalliamvakou2014promises,muse2022fixme}}: (i) less than one issue, (ii) the latest commit was not submitted within one year from the time spot of our data collection (i.e., September 2022), (iii) non-fork projects, and (iv) less than three contributors.
    \item \textbf{C3}: We then filter out those projects that do not contain significant build investments.
    As \citet{smith2011software} claimed, the issue of build system maintenance does not arise until a system reaches a certain level of maturity.
    In this step, we consider Ivy and Ant as a whole since Ivy is an extension of Ant.
    To identify build files, we refer to the list of filename conventions provided by~\citet{10.1145/2591062.2591181}.
    For example, the convention of Maven build specification files is \texttt{pom.xml and maven([123])?.xml}.
    Similarly, we draw a threshold plot to determine the lines of code and we filter out those projects whose lines of code are less than 500 \fix{approximating the point of diminishing returns}.
    \item \textbf{C4}: The last criterion is to further filter out those build investments that do not have significant lines of comments, since our study focuses on SATD comments. 
    Hence, we exclude build investments where the lines of comments are less than 60, relying on the threshold plot \fix{as a diagnostic tool to select a threshold approximating the point of diminishing returns}. The description of extracting comments is explained in Section~\ref{sec:com}. 
\end{itemize}
In the end, a total of 6,502 projects meet the above four criteria, i.e., 1,403 projects from Autotools, 3,453 projects from CMake, 1,535 projects from Maven, and 111 projects from Ant/Ivy, respectively.
Note that the threshold plots generated in C3 and C4 are available in our replication package~\citep{replicate} \fix{to prevent redundancy by displaying similar plots within
the paper}.

\subsection{SATD Comments Extraction}
\label{sec:com}
We invoke \texttt{ANTLR4} lexical analyzers for CMake and Autotools to process build files of our 6,502 projects that are collected in Section~\ref{sec:proj}. 
In addition, we also consider \texttt{dnl} M4 macro\footnote{\url{https://www.gnu.org/software/m4/manual/html_node/Dnl.html}} that discards all input that follows it on the same line as a comment in our study. 
For XML-based build systems (i.e., Ant, Ivy, and Maven), we utilize \texttt{ElementTree.XMLParser} libary\footnote{\url{https://docs.python.org/3/library/xml.etree.elementtree.html}} to extract comments that are recognized as content appearing between \texttt{<!––} and \texttt{––>} XML tokens. Finally, we retrieve 2,628,919 comments from build specification files in total. 
Similar to our prior work~\citep{9551792}, we identify SATD comments using the keywords-based approach of \citet{potdar2014exploratory}. 
We further discuss the rationale of electing the keywords-based approach instead of the existing machine learning based approach in Section~\ref{threats}.
We are able to identify 64,013 SATD comments in build specification files from 4,641 projects. Table~\ref{tab:satd} shows the distribution of obtained SATD and non-SATD comments in each build system.

To assess the accuracy of the keyword set used to identify SATD comments, we conduct a qualitative study of a statistically representative and stratified sample of comments that are not identified as SATD by the keyword-based approach. The sample size is calculated to achieve estimates with a 95\% confidence level and a 5\% confidence interval.
In total, we randomly select 384 comments from a pool of 2,564,906 comments that are not identified as SATD comments. 
The first two authors then manually examine these comments to determine whether or not they are SATD comments. 
To mitigate the threat of subjective nature, we calculate the Cohen's Kappa agreement of our codes between the first two authors, whereas Cohen’s Kappa
is 0.87, indicating an `Almost perfect' agreement~\citep{viera2005understanding}.
In the end, we find that only 21 comments (5.4\% of the total sample) are incorrectly missed by the keyword-based approach, suggesting that our employed approach is robust.

\subsection{SATD Clones Identification}
\label{sec:clone}
We apply a list of steps to identify groups of duplicate or near-duplicate SATD comments as SATD clones from 64,013 obtained SATD comments, by leveraging the method of \citet{10.1145/3524610.3528387}. We now describe the details of each step below.  

\textbf{CI1: Text \fix{Preprocessing}.}
To obtain more weights on each word in hyperlinks, we separate words in hyperlinks with space by using the following regular expression:
\path{
https?:\/\/(www\.)?[-a-zA-Z0-9@:%._\+~#=]{2,256}\.[a-z]{2,6}\b([-a-zA-Z0-9@:%_\+.~#?&//=]*)
}.
In the SATD comment extraction, we regard \texttt{dnl} M4 macro as one of the comment formats in the Autotools build system. Thus, we remove ``dnl'' if the comment starts with it.
Moreover, to reduce the impact of noisy text in comments, we remove special characters by using the regular expression \verb/[^A-Za-z0-9]+/. Since stop words (e.g., ``for'' and ``until'') could convey critical semantics in the context of SATD comments, we opt to exclude stop word removal~\citep{wait_for_it, 9551792}. We further filter out uninformative SATD comments that contain a single word (e.g., ``TODO'') since these annotations are highly likely to be cloned in the software development process.

\textbf{CI2: Feature Extraction.}
We apply SentenceTransformer~\citep{reimers-2019-sentence-bert}, a pre-trained BERT (Bidirectional Encoder Representations from Transformers) model, to generate a sentence embedding for each SATD comment. This model is broadly used to produce document vectors of SATD clones in source code comments and similar questions in Q\&A websites~\citep{10.1145/3524610.3528387, kamienski2023analyzing}. 

\textbf{CI3: Comment Similarity.}
We then calculate the pairwise cosine similarity from document vectors of SATD comments.
To do so, similar to the previous study~\citep{10.1145/3524610.3528387}, we exclude SATD comments that do not exist in any pair with a similarity score $\ge$ 0.8.


\textbf{CI4: Clustering.}
We employ the Density-Based Spatial Clustering of applications with Noise (DBSCAN) clustering algorithm~\citep{ester1996density} provided by the \texttt{Scikit-Learn} library~\citep{dbscan}. 
\minor{We set (i) the ``metric'' parameter to ``cosine'', which determines the cosine distance as the measure of similarity.
This metric captures the orientation-based similarity of data points, making it a more fitting choice than Euclidean distance for our analysis; (ii) an ``eps'' value of 0.1, which is a smaller value than SATD clones in source code comments~\citep{10.1145/3524610.3528387} to build clusters of a finer granularity. The ``eps'' parameter defines the maximum distance between two samples, as calculated by the chosen metric, for one to be considered within the neighborhood of the other; and (iii) ``min\_samples'' parameter to 2 to ensure that each cluster contains a minimum of two SATD comments.} 
We evaluate the clustering by silhouette score (1 to -1), which measures how well a set of samples is clustered and the separability between clusters~\citep{rousseeuw1987silhouettes}. 
We obtain a score of 0.96, indicating a small intra-cluster average distance. 
SATD comments that are not been clustered to any group are removed from the dataset.
Finally, we obtain 5,074 groups from 52,870 SATD comments.

\begin{figure}[t]
  \centering
  \includegraphics[width=.8\linewidth]{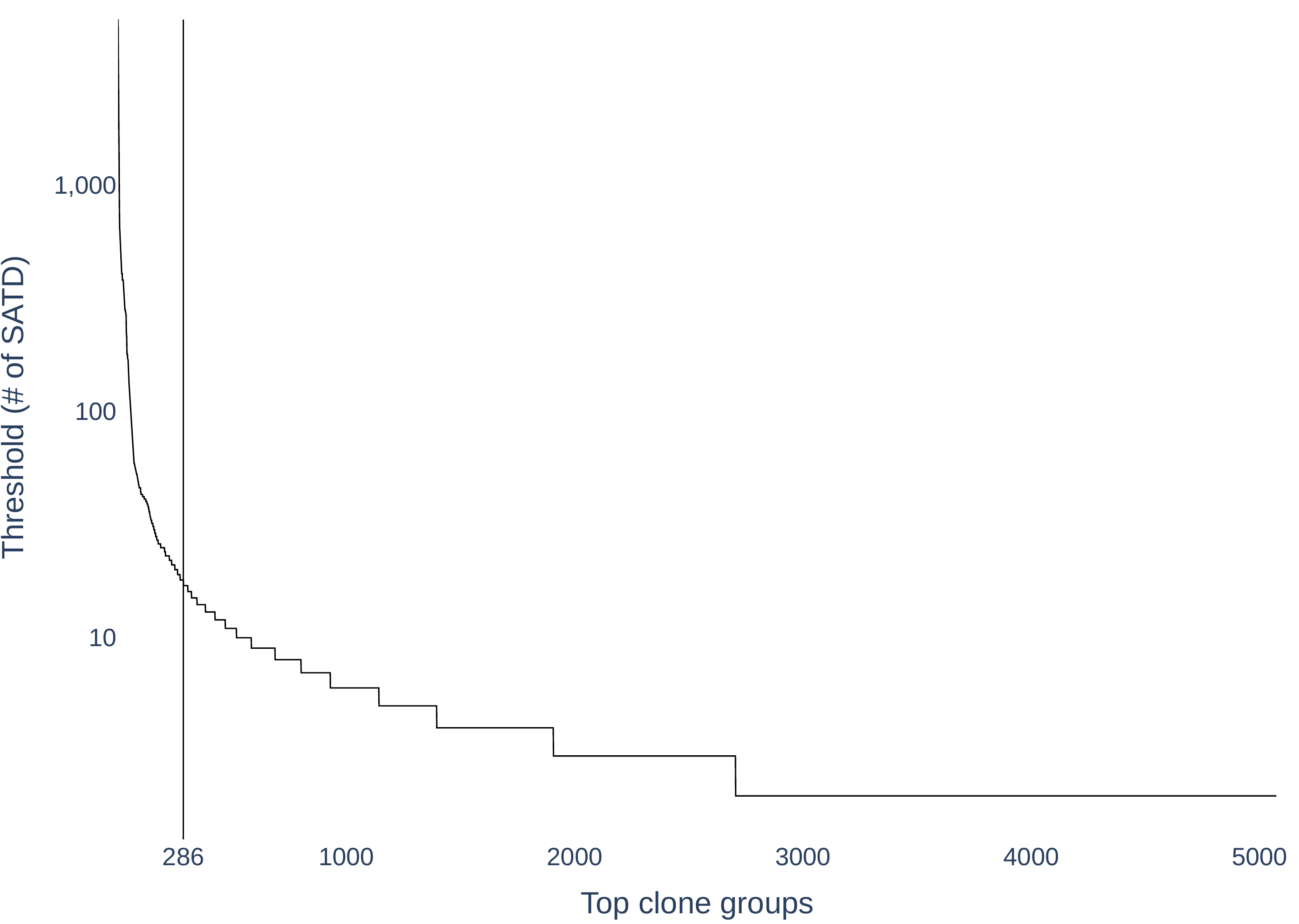}
  \caption{Threshold plot for top groups of SATD clones.}
  \label{fig:hl}
\end{figure}

\textbf{CI5: Human Labeling.}
To mitigate the effect of false positives from the keywords-based approach in the high-impact (pervasively cloned) SATD groups, we rank 5,074 groups by the number of SATD comments in each group. 
Specifically, we draw a threshold plot to determine the high-impact SATD groups, with a threshold of 286 groups \fix{that approximates the point of diminishing returns}, as shown in Figure~\ref{fig:hl}. 
The first two authors manually validate these comments (one representative comment in each group of SATD clones) to
determine whether or not they are SATD comments. We find that 29 SATD groups are false positives. Eventually, we obtain 5,045 groups consisting of 50,608 SATD comments, as shown in Table~\ref{tab:satd}.

\begin{table}[t]
\caption{Distributions of SATD and non-SATD comments.}
\label{tab:satd}
\centering
\resizebox{.99\columnwidth}{!}{
\begin{tabular}{llrrrrrrr}
\toprule
\multirow{5}{*}[-1.5em]{\rotatebox[origin=c]{90}{\textbf{SATD}}}                 &      & \textbf{Autotools} &\textbf{CMake} & \textbf{Maven} & \textbf{Ant} & \textbf{Ivy} & \textbf{Sum}   & \textbf{\# repo} \\
\midrule
&\# SATD     & 34,491     & 26,394   & 2,524   & 582   & 22  & 64,013   & 4,641 \\
&CI1  & 34,409     & 26,111 & 2,498  & 566 & 22  & 63,606 & 4,629    \\
&CI3 & 32,990     & 18,826 & 1,685  & 404 & 2   & 53,907 & 3,627    \\
&CI4        & 32,884     & 18,040 & 1,561  & 385 & 0   & 52,870 & 3,458    \\
&CI5                     & 30,972     & 17,712 & 1,561  & 363 & 0   & 50,608 & 3,427 \\
&\fix{cloning rate} & \fix{95\%} & \fix{68\%} & \fix{62\%} & \fix{65\%} & \fix{-} & \fix{-}  & \fix{-}\\
\midrule
\multirow{4}{*}[-0.1em]{\rotatebox[origin=c]{90}
{\textbf{non-SATD}}}& \# non-SATD &
1,230,817   & 1,039,442 & 242,689 & 51,643 & 315 & 2,564,906 &  6,502 \\
& \# sample                      & 67,574     & 41,447 & 3,465  & 935 & 43  & 113,464 & 4,588    \\
&CI1               & 53,742     & 39,617 & 3,124  & 862 & 41  & 97,386  & 4,549    \\
&CI3 & 51,509     & 29,402 & 1,985  & 560 & 5   & 83,461  & 4,055    \\
&CI4                    & 51,062     & 27,712 & 1,791  & 544 & 0   & 81,109  & 3,780   \\
&\fix{cloning rate} & \fix{76\%} & \fix{67\%} & \fix{52\%} & \fix{58\%} & \fix{-} & \fix{-}  & \fix{-}\\
\bottomrule
\end{tabular}}
\end{table}



\begin{figure}[t]
  \centering
  \includegraphics[width=.8\linewidth]{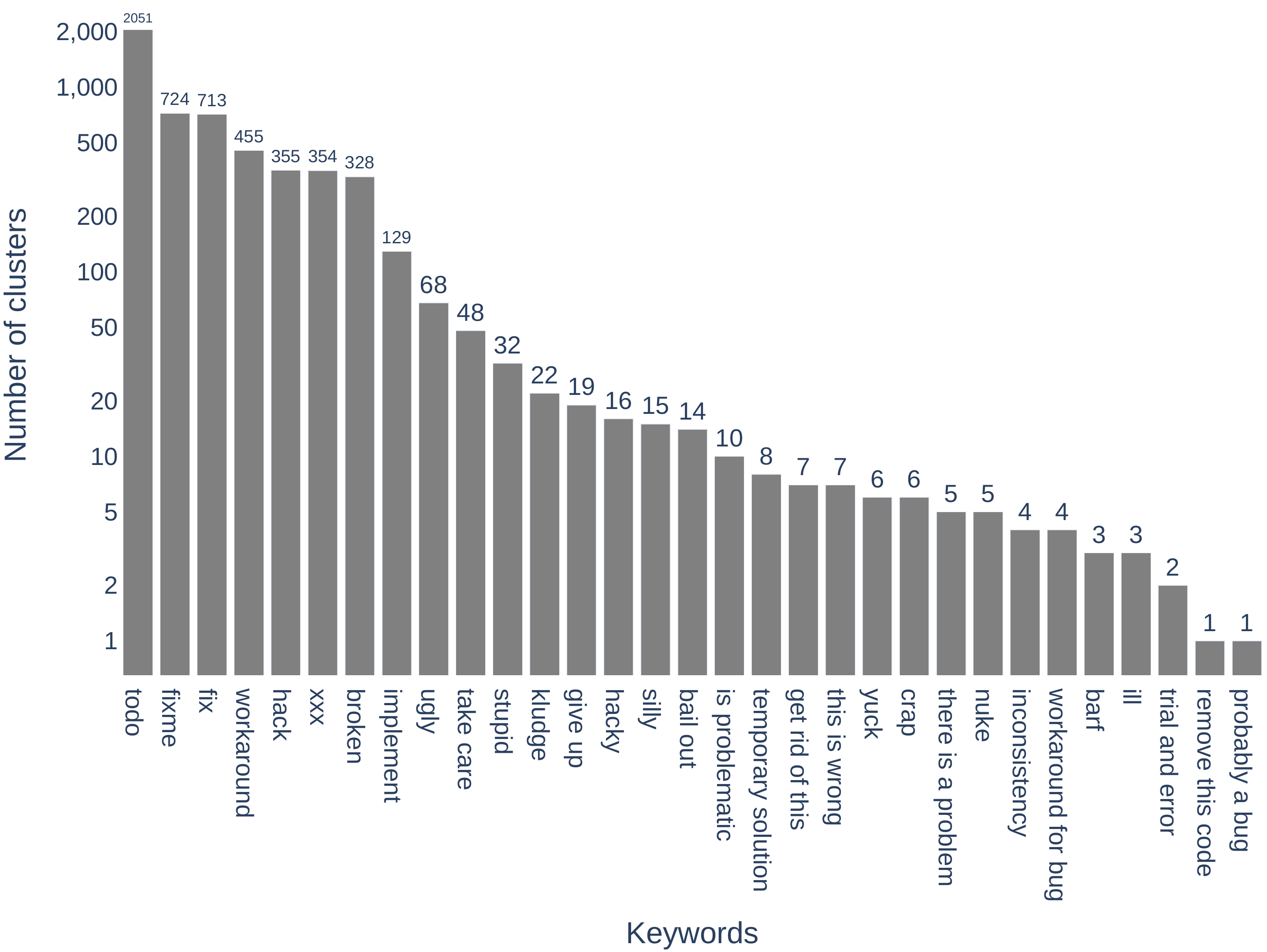}
  \caption{Number of clusters for each SATD keyword \minor{on the log scale}.}
  \label{fig:dis}
\end{figure}

We further performed an analysis to examine
how the SATD keywords are distributed over the clusters. Figure~\ref{fig:dis} presents the distribution of SATD keywords across the 5,045 groups. Notably, the `todo' keyword is prevalent, appearing in 2,051 SATD clone groups. Other indicators like `fixme', `fix', `workaround', `hack', `xxx', and `broken' are also prominently found across numerous clone groups, accounting for at least 328 groups.

\subsection{Non-SATD Clones Preparation}
It is largely unclear the prevalence of comment clones in build systems in general. 
Thus, we prepare a set of non-SATD clones as our \textbf{baseline} that is used in \textbf{RQ1--RQ3}, to gain a better understanding of the SATD clones via comparison. To do so, \minor{we adopt a focused approach by extracting two non-SATD comments adjacent to each SATD comment within the same build file (i.e., one from above the SATD comment and another from below the SATD comment in the same build file), yielding 113,464 non-SATD comments from 4,588 projects.  This selection strategy aims to investigate whether the practice of cloning non-SATD comments is consistent with that of SATD comments in similar program contexts during our subsequent analysis.  
As shown in Table~\ref{tab:satd},
7,368 groups of non-SATD clones consisting of 81,109 non-SATD comments are identified from 3,780 projects.}

\section{Empirical Results}
\label{emp}

In this section, we present the results for each of our research questions.

\subsection{Prevalence of SATD Clones (RQ1)}
\label{RQ1}
\fix{\RqOne}

\smallskip
\noindent
\textbf{\emph{\underline{Approach.}}}
To gain an understanding of the extent to which cloning is present in the SATD, we first conduct a quantitative analysis to explore the prevalence of SATD clones \fix{(i.e., 5,045 clone groups from 50,608 SATD
comments)}. 
The proportion of SATD clones is measured to quantify the prevalence. 
Since we removed false positives from SATD comments in the human labeling process (CI5 in Section~\ref{sec:clone}), the formula for calculating the proportion is defined as follows:
$\frac{\mathbf{S_{CI5}}}{ (\mathbf{S_{Orginal}} - (\mathbf{S_{CI4}} - \mathbf{S_{CI5}}))}$,
where $\mathbf{S_{CI5}}$ is the number of SATD comments after the human labeling process, $\mathbf{S_{Orginal}}$ represents the number of SATD comments before the text preprocessing process, and  $\mathbf{S_{CI4}}$ denotes the number of SATD comments after the clustering process.
Given an example, for Ant build system, $\mathbf{S_{Orginal}}$, $\mathbf{S_{CI4}}$, $\mathbf{S_{CI5}}$ are 582, 385, 363, separately as shown in Table~\ref{tab:satd}.

Inspired by our motivating example illustrated in Section~\ref{mot}, we further examine the SATD clones in the following three dimensions:
\textit{(I) repository} -- SATD comments are cloned in the internal project or across the external projects; 
\textit{(II) build tool} -- SATD comments are cloned in the same build tool or across the different build tools; 
and \textit{(III) programming language} -- SATD comments are cloned in the build tools that are adopted in projects which are
implemented using the same or different programming languages.

\smallskip
\noindent
\textbf{\emph{\underline{Results.}}}
Table~\ref{tab:rq1} shows the distribution of SATD clones and non-SATD clones in terms of three dimensions (repository, tool, and language) and Table~\ref{tab:system} further presents the distribution of clones within the same build tool. 

\ul{\textit{SATD comments are frequently cloned in build systems.}}
The results show that the proportions of the SATD clones are 95\%, 68\%, 62\%, 65\% for Autotools, CMake, Maven, and Ant, respectively. This suggests that SATD clones are frequently
occurring in build systems, accounting for at least 62\% of SATD comments.
In contrast, the proportions of the non-SATD clones are 76\%, 67\%, 52\%, 58\% for Autotools, CMake, Maven, and Ant, respectively.
Based on the above results, SATD comments are relatively more likely to be cloned in build systems, when compared to non-SATD ones. Moreover, we find that the proportions of SATD clones are higher among build systems than those reported in source code (41--65\%)~\citep{10.1145/3524610.3528387}.

\begin{table}[t]
\caption{The distribution of SATD clones and non-SATD clones in dimensions of repository, tool, and language.}
\label{tab:rq1}
\centering
\resizebox{\columnwidth}{!}{
\begin{tabular}{llr@{}rr@{}rrrr}
\toprule
                  &    & \multicolumn{2}{c}{\# Groups} & \multicolumn{2}{c}{\# SATD   } & Mean      & Median & Maximum       \\
\midrule
\multirow{6}{*}[-1.5em]{\rotatebox[origin=c]{90}{SATD}} & Internal repository clone & 1,242 & (25\%)   & 4,030 & \fix{(8\%)}     & 3.24  & 2      & 56      \\
&External repository clone & 3,803 & (75\%)     & 46,578 & \fix{(92\%)}   & 12.25 & 3      & 5,413                         \\
\cmidrule{2-9}
&Same language clone       & 5,026 & (99\%)    & 50,302 & \fix{(99\%)}   & 10.01 & 3      & 5,413                           \\
&Cross-language clone      & 19  & (1\%)     & 306 & \fix{(1\%)}    & 16.11 & 7      & 93                             \\
\cmidrule{2-9}
&Same tool clone         & 5,000 & (99\%)    & 50,012 & \fix{(99\%)}  & 10.00 & 3      & 5,413                           \\
&Cross-tool clone        & 45   & (1\%)    & 596 & \fix{(1\%)}    & 13.24 & 6      & 93       \\
\midrule
\multirow{6}{*}[-1.5em]{\rotatebox[origin=c]{90}{non-SATD}} & Internal repository clone & 1,211  & (16\%)  & 4,455 & \fix{(5\%)}    & 3.68  & 2      & 84       \\
&External repository clone & 6,157 & (84\%)     & 76,654 & \fix{(95\%)}   & 12.45 & 3      & 2,427     \\
\cmidrule{2-9}
&Same language clone       & 7,332  & (99\%)   & 78,769 & \fix{(97\%)}   & 10.74 & 3      & 2,427           \\
&Cross-language clone      & 36  & (1\%)     & 2,340 & \fix{(3\%)}    & 65 & 7.5    & 598                        \\
\cmidrule{2-9}
&Same tool clone         & 7,249 & (99\%)    & 77,316 & \fix{(95\%)}   & 10.67 & 3      & 2,427                     \\
&Cross-tool clone        & 119  & (1\%)    & 3,793  & \fix{(5\%)}   & 31.87 & 6      & 598   \\
\bottomrule
\end{tabular}}
\end{table}

\ul{\textit{Dominate SATD comments are propagated under the same build tool and most of them are found to be cloned in external repositories.}} 
As we can see from Table~\ref{tab:rq1}, in terms of repository dimension, 25\% of groups of SATD clones are from the internal repository while 75\% of them are from the external repository.
In terms of language and build tool dimensions, we observe that the groups of SATD clones are almost from the same language and the same build tool, with 99\% being identified for both dimensions. 
Compared to non-SATD clones, similar observations are found regarding the language and build tool dimensions (i.e., 99\% for both). 
However, the proportion of the non-SATD clones from the external repository is greater than the SATD ones (84\% against 75\%). \fix{The maximum cloning frequency for SATD clones, at 5,413 instances, is nearly twice that of non-SATD clones, which stands at 2,427 instances. A detailed discussion on the most prevalently cloned SATD comments can be found in Section~\ref{mot}.}




\begin{table}[t]
\caption{The distribution of \fix{5,000 SATD} same tool clones and \fix{7,249 non-SATD same tool clones in the tools}.}
\label{tab:system}
\centering
\begin{tabular}{llr@{}rr@{}rrr}
\toprule
& & \multicolumn{2}{c}{\# Groups} & \multicolumn{2}{c}{\# SATD} & Mean & Median \\
\midrule
\multirow{4}{*}{\rotatebox[origin=c]{90}{SATD}} & Autotools & 1,045 & \fix{(21\%)} & 30,732 & \fix{(61\%)} & 29.41 & 4 \\
& CMake & 3,541 & \fix{(71\%)} & 17,439 & \fix{(35\%)} & 4.92 & 3 \\
& Maven & 349 & \fix{(7\%)} & 1,492  & \fix{(3\%)}& 4.28 & 2 \\
& Ant & 65 & \fix{(1\%)} & 349 & \fix{(1\%)} & 5.37 & 3 \\
\midrule
\multirow{4}{*}{\rotatebox[origin=c]{90}{non-SATD}} & Autotools & 1,962 & \fix{(27\%)} & 49,955 & \fix{(65\%)} & 25.46 & 4 \\
& CMake & 4,889 & \fix{(67\%)} & 25,811 & \fix{(33\%)} & 5.28 & 3 \\
& Maven & 325 & \fix{(4\%)} & 1,178 & \fix{(2\%)} & 3.62 & 2 \\
& Ant & 73 & \fix{(1\%)} & 372 & \fix{(1\%)} & 5.10 & 2 \\
\bottomrule
\end{tabular}
\end{table}

\ul{\textit{Clones of SATD comments in Autotools are the most pervasive.}}
\fix{Be a} closer look at the clones within the same build tool, as shown in Table~\ref{tab:system}, we observe that SATD comments in Autotools are pervasively cloned (i.e., 30,732 SATD comments emerged into 1,045 groups) when compared to the other three studied build tools.
For example, the mean and the median values per group of SATD clones in Autotools are 29.41 and 4. 
While the mean and the median values of the other three build tools (i.e., CMake, Maven, and Ant) vary from 4.28 to 5.37 and from 2 to 3, separately.
We also find that non-SATD comments in build systems exhibit a similar pattern in cloning, that is, the mean and median values of Autotools are 25.46 and 4, accounting for the greatest.

\fix{The differences between SATD clones and non-SATD clones are: (i) SATD comments exhibit a higher likelihood of being cloned in build systems, ranging from 62\% to 95\%, in comparison \fix{to non-SATD comments, which range from 52\% to 76\%; (ii) the peak cloning frequency for SATD clones is 5,413 instances, almost double that of non-SATD clones, which have a frequency of 2,427 instances; and (iii) SATD comments are more frequently propagated within the internal repository, with a proportion of 25\%, as opposed to non-SATD comments at 16\%. Conversely, non-SATD comments see a broader propagation in external contexts, standing at 84\%, compared to SATD comments at 75\%.}}

\fix{\ul{\textit{More than half of SATD clones across build tools are annotation templates.}}
The results shown in Table~\ref{tab:rq1} reveal that SATD clones can be found across build tools, i.e., 596 SATD comments from 45 groups.
Note that 19 of them (containing 306 SATD comments) are cloned in build tools that are adopted in projects that are implemented using different programming languages, referred to as cross-language clones. 
To gain a better insight into this, we manually inspected 596 SATD comments across build tools to deduce the reasons. 
We also analyzed 3,793 non-SATD comments from 119 groups for comparison.}


\fix{Table~\ref{tab:cross} presents the reasons for SATD clones across build tools (CTC) and programming languages (CLC). We find that more than half of SATD clones are annotation templates. Although we filtered out many of these uninformative SATD comments by removing SATD comments that contain single words in the text preprocessing step of Section~\ref{sec:clone}, some examples permeate through to our dataset. For example, developers add comments that ask for clarification/rationale (e.g., ``\# todo is this correct/needed?''), refactoring in the future (e.g., ``\# TODO: Remove this when fixed''), and store version information (e.g., ``\# TODO check version compatibility''). We also observe that 34\% of SATD comments are annotation closure, e.g., ``End Fixme'' or ``End of workaround''. We find that 16 SATD comments were cloned due to workarounds for different issue reports on \texttt{Jira}. Most issue reports in SATD comments are annotated in Java build systems. However, we find one of those issue reports on \texttt{Jira} is also annotated in CMake.\footnote{\url{https://github.com/v6d-io/v6d/blob/9c6bb0c2cb31c54d6e798555d5a92cb067490083/CMakeLists.txt\#L263}}
For SATD clones that only span across build tools (CTC), we observe that 43\% of SATD comments are specific issues that are shared in Autotools and CMake. These issues include an \texttt{XCode 11} bug, a broken implementation of \texttt{log1p} function in \texttt{OpenBSD}, a header file (i.e., \texttt{ifaddrs.h}) that is missing in particular deployment contexts, and a scenario where the \texttt{Sun linker} is incorrectly being used when \texttt{GCC} should have been. 
We observe that 29\% of SATD comments are annotation templates, and seven SATD comments contain workarounds for issue reports on \texttt{Ubuntu} packages. Additionally, we find that 85 SATD comments (23\% of SATD comments) annotate one common release
setting\footnote{\url{https://github.com/turican0/remc2/blob/92e8b255b696fdfb4942324bab5e5f092c4ed69f/sdl/configure.in\#L8}} 
across CMake and Autotools build systems from 38 GitHub repositories.}

\fix{Among the non-SATD comments across build tools, the primary reason for both CTC and CLC is the use of standard commenting practices in build systems, accounting for 83\% and 7\%, respectively. 
These practices include commented-out Autoconf macros and bash script headers. 
The remaining non-SATD clones are related to copyright, with 17\% and 93\% of instances being classified for CTC and CLC, separately.
Note that SATD comments are identified using specific keywords, while non-SATD comments arise from the absence of these keywords, making it challenging to discern a consistent pattern in non-SATD clones.
The details of our coding results are included in our replication package~\citep{replicate}.}

\begin{table}[t]
\caption{Reasons for SATD clones across build tools and programming languages.}
\centering
\label{tab:cross}
\resizebox{.8\columnwidth}{!}{
\begin{tabular}{llr@{}r}
\toprule
& \textbf{Reason} & \multicolumn{2}{c}{\textbf{Frequency}} \\
\midrule
\multirow{3}{*}[-1.0em]{\rotatebox[origin=c]{90}{ CLC}} & Annotation template   & 187 & (61\%)      \\
& Annotation closure & 103 & (34\%)      \\  & Workaround for issue reports on Jira           & 16 & (5\%) \\
\cmidrule{2-4}
& sum & 306 & 100\% \\
\midrule
\multirow{4}{*}[-1.0em]{\rotatebox[origin=c]{90}{CTC}}  & Specific issues shared in Autotools and CMake & 126  &  (43\%) \\ & Annotation template  & 85 & (29\%)  \\ & Release setting  & 70  & (23\%)     \\ & Workaround for issue reports on Ubuntu packages & 7    & (2\%)    \\ & False positive  & 2  & (1\%) \\
\cmidrule{2-4}
& sum & 290 & 100\% \\
\bottomrule
\end{tabular}}
\end{table}

\begin{tcolorbox}
\textbf{RQ1 Summary:}
SATD comments are frequently cloned in build systems and most of them are found to be cloned in the external repository (75\%). 
Particularly, clones of SATD comments in Autotools are the most pervasive. \fix{The predominant factor driving the cloning of SATD comments across build tools can be attributed to the introduction of annotation templates specific to various build systems. }
\end{tcolorbox}

\subsection{Prevalence of Statement Clones (RQ2)}
\label{RQ2}
\fix{\RqTwo}

\smallskip
\noindent
\textbf{\emph{\underline{Approach.}}}
In this research question, we further explore the prevalence of statement clones that surround SATD clones as we speculate that SATD clones are likely to be introduced due to the similar statements surrounding them. 
Specifically, we extract upper and lower blocks of statements in a threshold of five lines and calculate their similarity within a group of SATD clones. \fix{This threshold has been widely used in previous studies, as highlighted by~\citet{10.1145/2591062.2591181} and~\citet{10.1145/3524610.3528387}. For instance, the benchmark study on build statement clones
demonstrated that a threshold ranging between five and nine lines yielded optimal
clone coverage results~\citep{10.1145/2591062.2591181}. To further address the threat of threshold in different context, we conducted a statistical test to confirm the impact of the number of lines.
Specifically, we performed a one-way ANOVA test~\citep{fisher1970statistical} to compare thresholds of 5, 10, 15, and 20 lines. 
The results suggested that there is no significant difference between these thresholds (with \textit{F = 1.941} and \textit{p = 0.121} for mean values of similarity scores in SATD clone groups, as well as \textit{F = 2.3} and \textit{p = 0.0752} for median values of similarity scores in SATD clone groups). 
Hence, based on these findings and to minimize computational costs, we elected to set our threshold at five lines.}
Our interest is in statement clones in terms of different build systems, hence we use clones from same tool clones (5,000 groups comprised of 50,012 SATD comments) from \textbf{RQ1} as our studied datasets in this RQ. 

\textit{Similarity calculation of statements}.
We leverage the method provided by~\citet{hong2022commentfinder} and the details are described as follows:
First, we remove punctuation characters to ensure that the code tokens will not be artificially repetitive. Since statements are case-sensitive similar to source code, we opt to exclude performing lowercase and reduction of inflectional forms (i.e., stemming and lemmatization). Then, 
we represent statements as vectors using the Bag-of-Words model with the \texttt{Countvectorize} function of the \texttt{Scikit-Learn} library~\citep{countvect}. \fix{Similar to \citet{hong2022commentfinder}}, we remove code tokens that commonly appear across the statements (i.e., tokens that appear more than 50\% of the statements).
We then apply cosine similarity to measure statement clones in each group of SATD clones. 

\textit{Comparison against baseline}.
For the baseline (non-SATD clones), we adopt the same method to calculate the similarity scores of their statement clones.
Furthermore, we perform a Mann-Whitney test~\citep{mann1947},  a non-parametric statistical test, to examine any significant difference in similarity scores of statement clones between SATD and non-SATD with $\alpha$ =0.05.
We also measure the effect size using Cliff's $\delta$, a non-parametric effect size measure \citep{Cliff:1993}.
Effect size is analyzed as follows: 
(1) $|\delta| < 0.147$ as Negligible, (2) $0.147 \leq |\delta| <0.33$ as Small, (3) $0.33 \leq |\delta| <0.474$ as Medium, or  (4) $0.474 \leq | \delta|$ as Large \citep{romano:2006}.

\begin{figure}[t]
  \centering
  \includegraphics[width=.8\linewidth]{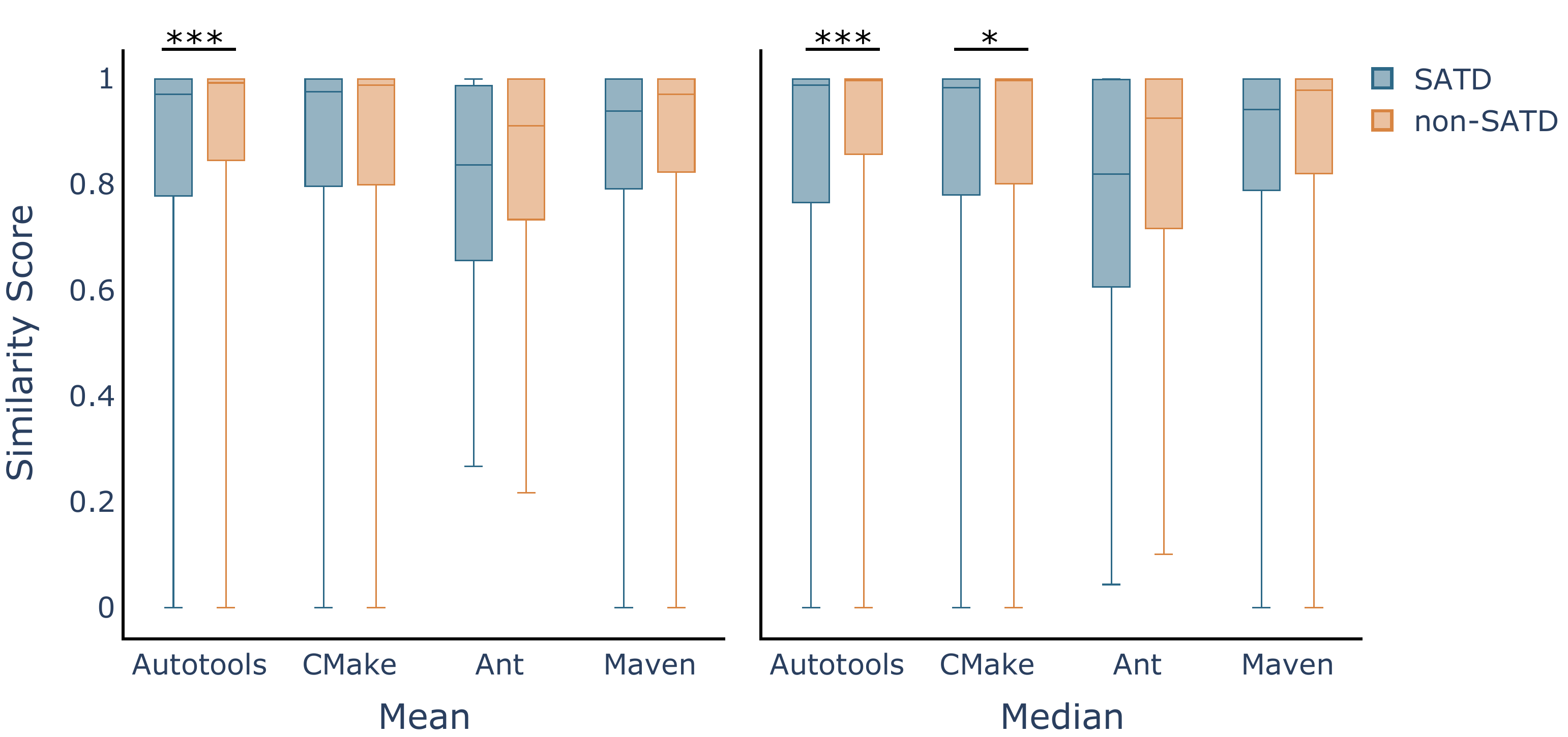}
  \caption{Mean and median of similarity scores of \fix{statements surrounding each comment in SATD/non-SATD group} (* p-value $<$ 0.05; ** p-value $<$ 0.01; and *** p-value $<$ 0.001).}
  \label{fig:rq2}
\end{figure}

\smallskip
\noindent
\textbf{\emph{\underline{Results.}}}
Figure~\ref{fig:rq2} presents the distribution of the mean and median values of the similarity scores of statements between SATD and non-SATD clones.

\ul{\textit{Similarity scores of statements surrounding SATD clones are relatively lower compared to non-SATD ones.}}
As shown in the figure, we can see that statements surrounding SATD clones are highly similar. 
Most of the similarity scores of the statement clones are greater than 0.8 (i.e., with \fix{73}\% of groups of SATD clones having a median similarity score, and \fix{74}\% of them having a mean similarity score), suggesting that they tend to be duplicate or near duplicate. 
This finding indicates that SATD clones may be introduced due to the similar statements surrounding them.
Surprisingly, we observe that the similarity scores of statements surrounding non-SATD clones are relatively higher than SATD clones no matter the median or the mean of the scores. \fix{From the figure, it is evident that the statements around comment clones (both SATD and non-SATD) vary across different build systems. While the third quartile for all build systems reaches a similarity score of 1 (indicating identical statements), the first quartile for the Maven build system surpasses the others. This higher similarity in Maven may be attributed to the tag representation inherent to its XML structure, leading to greater similarities in build statements. }

Furthermore, the Mann-Whitney test confirms that statements surrounding SATD and non-SATD clones in Autotools have significant differences (i.e., \textit{p-value} $<$ 0.05) with negligible effect size in terms of mean \fix{(i.e., $|$\textit{$\delta$}$|$ = 0.0983 $<$  0.147) and median (i.e., $|$\textit{$\delta$}$|$ = 0.0973 $<$  0.147)} values. 
Such a significant difference \fix{is also found} in CMake with a negligible effect size \fix{(i.e., $|$\textit{$\delta$}$|$ = 0.0293 $<$ 0.147)} in terms of median values. 
This result indicates that although the statements surrounding SATD clones are less likely to be similar than the ones surrounding non-SATD clones significantly, the difference is marginal. \minor{Therefore, we found either no clear evidence to support or only a small effect size to reject our hypothesis that SATD clones are likely to occur due to cloned statements surrounding them.}

\begin{tcolorbox}
\textbf{RQ2 Summary:}
The statements surrounding SATD clones are highly similar. Compared to those surrounding non-SATD clones, the similarity scores of SATD clones are relatively lower than non-SATD ones.
\end{tcolorbox}

    

\subsection{Authorship of SATD Clones (RQ3)}
\fix{\RqThree}

\smallskip
\noindent
\textbf{\emph{\underline{Approach.}}}
In \textbf{RQ3}, we aim to explore the authorship of SATD clones in build systems. 
We assume that the majority of SATD clones are introduced by the author of the original SATD statements. 
Since our focus is on the SATD authorship across repositories, we use those SATD comments that are cloned across the external repository from \textbf{RQ1} (3,803 groups consisting of 46,578 SATD comments) as our studied datasets in this RQ.

\textit{Authorship identification}.
We use the \texttt{git log} command to extract the \texttt{git} history of SATD comments, i.e., \texttt{git log ----pretty=format:\%H,\%an,\%ae,\%at,\%B ----no-patch -L start,end:filename}. This command can obtain the metadata (introduced commit SHAs, author name, author email, introduced timestamp, and commit message) of a code with a line range. We notice that the introduced commit SHAs of SATD clones could also be duplicated in the different repositories (see more detail in the discussion section). Thus, we remove duplicate SATD comments in each group of SATD clones by the repository name (e.g., owner/name) and the commit SHA. 

\textit{Comparison \fix{against} baseline}.
Similarly, we identified the authorship of non-SATD clones using the above method.
We then introduce the following two related metrics to compare:

\ding{114} \textbf{Unique Author Density (UAD)}: $\mathbf{UAD}=\frac{\mathbf{Unique(authors)}}{ \mathbf{S_{group}} }$, where $\mathbf{Unique(authors)}$ refers to the number of unique authors (identical by author name and author email) in each group of clones and $\mathbf{S_{group}}$ represents the number of SATD comments in each group of SATD clones.

\ding{114} \textbf{Maximum Clone Density (MCD)}: $\mathbf{MCD}=\frac{\mathbf{Max(authors)}}{ \mathbf{S_{group}} }$, where $\mathbf{Max(authors)}$ denotes the maximum number of times of a SATD has been cloned by a single author and $\mathbf{S_{group}}$ is the number of SATD comments in each group of SATD clones.

Similar to \textbf{RQ2}, we invoke the Mann-Whitney test and Cliff's $\delta$ to measure the significant difference between SATD clones and non-SATD clones and its effect size.

\fix{Moreover, we further analyze (i) the time interval, measured in the number of commits same as~\citep{bavota2016large}, from the first commit introduced SATD to the HEAD commit; (ii) the experience level of developers who introduced SATD comments, gauging their experience by the number of their prior commits; and (iii) the commit messages of the SHAs that introduce SATD clones. We compare these results for the significant difference between SATD clones and non-SATD clones by the Mann-Whitney test and Cliff's $\delta$.}

\begin{figure}[t]
  \centering
  \includegraphics[width=.8\linewidth]{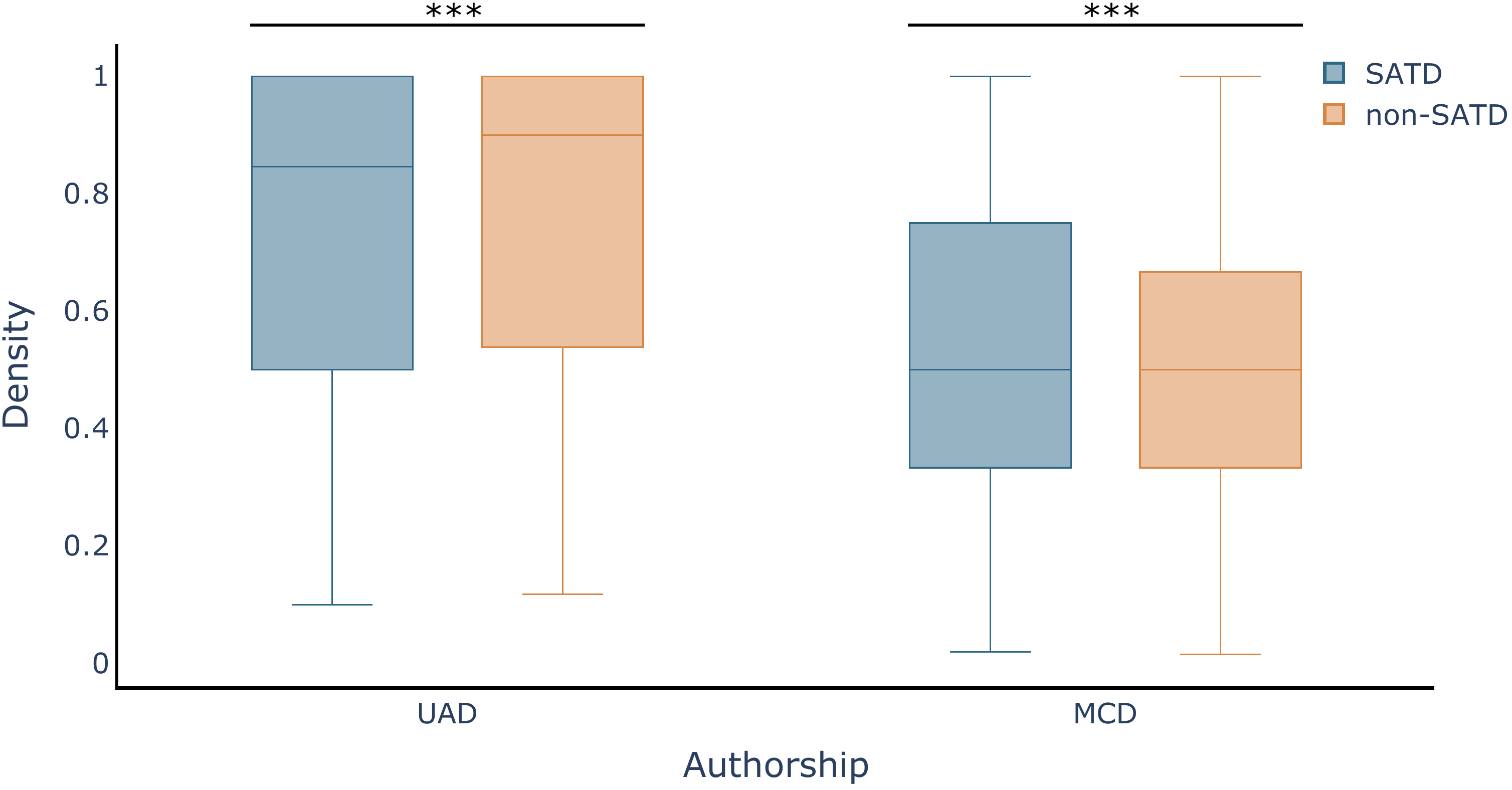}
  \caption{Authorship metrics of Unique Author Density (UAD) and Maximum Clone Density (MCD) between SATD/non-SATD clones (* p-value $<$ 0.05; ** p-value $<$ 0.01; and *** p-value $<$ 0.001).}
  \label{fig:rq3}
\end{figure}

\smallskip
\noindent
\textbf{\emph{\underline{Results.}}}
Figure~\ref{fig:rq3} presents the comparison results in terms of two defined metrics between SATD clones and non-SATD clones.

\ul{\textit{The majority of SATD clones are introduced by more than one author.}}
The statistical result shows that the number of groups of SATD clones that are introduced by the author of the original SATD statements is \fix{857}, accounting for only 23\% of groups of SATD clones from external repositories. 
This finding suggests that most of the SATD clones are introduced by more than one author.
In contrast, the number of groups of non-SATD clones that are introduced by the author of the original SATD statements is \fix{1,129}, accounting for \fix{18}\% of groups of non-SATD clones from external repositories. 

\fix{\ul{\textit{SATD comments tend to be cloned by relatively fewer authors compared to non-SATD ones.}}} 
For the Unique Author Density (UAD) metric, 
as shown in Figure~\ref{fig:rq3}, we observe that the median and first quartile value of non-SATD clones are slightly larger than SATD clones, indicating that cloned SATD comments are less diversely introduced by a unique set of authors. 
Moreover, a significant difference is observed between SATD and non-SATD via a statistical test, with negligible effect size \fix{(i.e., the \textit{$\delta$} value of 0.0558)}.
In terms of the Maximum Clone Density (MCD) metric, the third quartile value of SATD clones is greater than non-SATD ones, suggesting that a single author could clone more SATD than non-SATD. 
Furthermore, the Mann-Whitney test confirms a significant difference between SATD and non-SATD in MCD, with a negligible effect size (i.e., the \textit{$\delta$} value of \fix{0.0829}).

\fix{\ul{\textit{Developers often clone legacy SATD comments following the trend of comment clones in build systems in general.}}
For the time interval between the first commit introduced SATD
to the HEAD commit (results are available in our replication package),  it shows 
 that median values of the interval between the initial and HEAD commit are often long for each SATD clone group. Indeed, the studied examples had
a median of 1,548 commits. Interestingly, median values of the interval between the initial and HEAD commit are also long for each non-SATD clone group. A Mann-Whitney test suggests that there is no significant difference between the clone intervals of SATD and non-SATD comments (i.e., the \textit{p-value} value of 0.7827).} 

\fix{\ul{\textit{Experienced developers tend to introduce SATD comments.}} For the experience level of developers who introduced SATD comments, we found the median number of prior commits for authors introducing SATD comments within each clone group stood at 134.5 (available in our replication package). Notably, authors of SATD comments seem to have more experience compared to those introducing non-SATD comments, with the latter having a median of 112 commits. A Mann-Whitney test indicates a significant difference in the experience levels of authors between SATD and non-SATD comments, as reflected by a negligible effect size (i.e., \textit{$\delta$} value of 0.0651). An extended analysis of the commit messages of the SHAs that introduce SATD clones and non-SATD clones (available in our replication package) shows that similarity scores of obtained commit messages in terms of mean and median values range from 0.2 to 0.6. This indicates that comment clones contain considerable degrees of both overlapping and unique information in commit messages. We observe that SATD and non-SATD comments have significant differences with negligible effect sizes in terms of mean and median values (\textit{$\delta$} values of 0.7701 and 0.0669, respectively).}

\begin{tcolorbox}
\textbf{RQ3 Summary:}
\fix{Nearly} a quarter of SATD clones are introduced by the author of the original SATD
statements. On the one hand, we observe that the number of unique authors of non-SATD clones is relatively larger than that of SATD clones. On the other hand, the maximum number of times that a single author has cloned SATD is relatively greater than that of non-SATD clones. 
\end{tcolorbox}

\subsection{Characterizing SATD Clones (RQ4)}
\fix{\RqFour}

\smallskip
\noindent
\textbf{\emph{\underline{Approach.}}}
To answer this RQ, we conduct a qualitative analysis to investigate the characteristics (locations, reasons, and purposes) of SATD clones that exist in build systems.
In particular, we perform manual coding on those pervasively cloned SATD comments. 
We especially rank 5,000 groups descendingly from \textbf{RQ1} (clones within the same tool) based on the number of SATD comments in each group of SATD clones. 
One representative comment in each group of SATD clones is selected to code. 
Since we believe that pervasively SATD clones are more impactful and significant, we elect to adopt the code saturation method similar to the prior work~\citep{hirao_fse,9551792}.

\textit{Coding Guidelines.} \fix{The categorization of (SA)TD in source code has been widely explored in the literature. \citet{alves2014towards} identified 13 types of TD by their root causes, gathering definitions and indicators from the literature, such as Architecture Debt, Build Debt, and Code Debt. Later, \citet{maldonado2015detecting} expanded these classifications to encompass SATD, segregating SATD comments in source code into five categories: design, defect, documentation, requirement, and test SATD. Similarly, \citet{bavota2016large} adapted TD types to SATD based on underlying causes, further distilling these causes into more nuanced categories (e.g., subdividing Defect Debt into known defects to fix and partially fixed defects).}

\fix{Different from those works on characterizing SATD in source code based on the general causes in the software development, we delve into more specific characteristics (location, reason, and purpose) of SATD clones and establish a broader theory of SATD in the context of build systems. Therefore, we leverage the combination of the prior two works~\citep{9551792, nejaticode} 
as our heuristic reference.
Our prior work~\citep{9551792} characterized the locations, reasons, and purposes of SATD comments from only the Maven build system. 
\citet{nejaticode} investigated the code reviews of build system specifications and proposed a taxonomy of issue patterns in the build files covering Maven, QMake, and CMake. 
Although there are existing two similar guidelines, we assume that cases where SATD comments are cloned would vary from the diverse build systems and thus the exact taxonomy that fits our study is not constructed yet.}


\textit{Code saturation}. 
We initiated a set of codes within the first 50 comments.
Same to our prior work~\citep{9551792}, we strive for theoretical
saturation~\citep{eisenhardt1989building} to achieve analytical generalization. 
Then, we set our saturation criterion to 50, i.e.,
the first two authors continued to code selected SATD comments until no new codes have been discovered for 50 consecutive comments. 
Then, they opened discussions on classifying SATD comments in terms of locations, reasons, and purposes and tried to reach a consensus on disagreements between them. 
During these discussions, the fourth author who has more than ten years of research experience on build systems,
resolved each disagreement and suggested possible improvements on the categories. 
For any changes in coding guides, we revised coded SATD comments to apply these changes. Finally, we reached saturation after coding 200 SATD comments. 
Since codes that
emerged late in the process may apply to earlier reviews, we performed another pass over all of the comments to correct miscoded entries and tested the level of agreement of our constructed codes.
In total, these coding sessions took approximately 25 hours.

\begin{table*}[t]
\centering
\caption{Definition and frequency of locations, reasons, and purposes}
\label{tab:code}
\resizebox{\columnwidth}{!}{
\begin{tabular}{lp{43mm}p{100mm}r@{}r}
\toprule
& \textbf{Category} & \textbf{Definition}  & \multicolumn{2}{c}{\textbf{Frequency}} \\
\midrule
\multirow{14}{*}{\rotatebox{90}{Location}} & \textbf{Externals} & \multirow[t]{5}{\linewidth}{Build code that configures factors that are external to the projects. These factors include platforms both hardware and software, tools that are used in the build execution, build plugin features/libraries/dependencies for building the project, and version of any of these artifacts (including build systems). } & \textbf{115} & \textbf{(58\%)}\\
& Platform configuration & 
& 45 & (23\%)   \\
&Tool configuration &
& 30 & (15\%)     \\
& Libraries and plugins &
& 28 & (14\%)    \\
& Artifact versioning &
& 12 & (6\%)    \\
\cmidrule{2-5}
 & \textbf{Behavioural} & \multirow[t]{6}{\linewidth}{Build code that configures the behavior of the build system. Such as environmental variables and flags that alter the commands, build variables or override inherited variables that are used in the build execution, project descriptive information, the configuration of avoids redundancies or duplicate configurations through inheritance, passive behavior that does not affect artifacts produced by the build system.} & \textbf{61} & \textbf{(30\%)}\\
&Dynamic settings &
& 35 & (18\%)      \\
&Build variables &
& 20 & (10\%)      \\
&Project metadata &
& 3 & (1\%)    \\
&Multi-directory configuration &
& 2 & (1\%)   \\ 
&Logging &
& 1 & (1\%)       \\
 \cmidrule{2-5}
 & \textbf{File System} & \multirow[t]{3}{\linewidth}{Build code that specifics file system on a logical or physical layer.} & \textbf{24} & \textbf{12}\% \\
&Logical file system & & 15 & (8\%)\\
&Physical file system &  & 9 & (5\%) \\
\midrule
\multirow{22}{*}{\rotatebox{90}{Reason}} & \textbf{Limitation} & \multirow[t]{3}{\linewidth}{Constraints imposed by the design or implementation of third-party libraries or development tools.} & \textbf{75} & \textbf{(38\%)}      \\  
& External tool limitation &           & 32 &(16\%)  \\
& External library limitation &            & 22 & (11\%)      \\ 
& Build tool limitation  &              & 21 &(11\%)  \\
\cmidrule{2-5}
& \textbf{Configuration} & \multirow[t]{3}{\linewidth}{Configuration issues during the compilation and build process, such as compiler configuration, symbol visibility, file path styles in different platforms, and checking existence of files/features/libraries/dependencies.} & \textbf{74} & \textbf{(37\%)}  \\
& Compiler configuration &  & 31 & (16\%)       \\ 
& Symbol visibility & & 22 & (11\%)       \\ 
&  Platform-specific setting &  & 11 & (6\%)\\
&  Feature existence & & 10 & (5\%) \\
\cmidrule{2-5} 
& \textbf{Dependency} & \multirow[t]{3}{\linewidth}{Dependency issues due to unavailable artifacts or assets, such as missing stale dependencies, management of internal dependencies, and dependency conflicts.}& \textbf{13} & \textbf{(7\%)}      \\ 
& Missing dependency  &  & 7 &(4\%)      \\
& Internal dependency management & & 5 &(3\%)      \\
& Dependency conflict & & 1 & (1\%) \\
\cmidrule{2-5} 
& \textbf{Code smell} & Violations of fundamentals of design principles, i.e., instances of poor coding practice in build files.& \textbf{12} & \textbf{(6\%)}      \\ 
\cmidrule{2-5} 
& \textbf{Recursive call} &  Coherence issues, recursive calls to invoke another build file. & \textbf{6} & \textbf{(3\%)}      
\\
\cmidrule{2-5}
& \textbf{Document} &\multirow[t]{2}{\linewidth}{Inadequate project description issues, such as licensing and metadata specification.} & \textbf{5} & \textbf{(3\%)}       \\
& Specify metadata  &  & 4 & (2\%)       \\
& Licensing  & & 1 & (1\%)       \\
\cmidrule{2-5} 
 & \textbf{Release and install behaviors} & \multirow[t]{3}{\linewidth}{Sanitize project before releasing or post-install files or program after building} & \textbf{3} & \textbf{(2\%)} \\
&Release & & 2 & (1\%)\\
& Post-install & & 1 & (1\%) \\
\cmidrule{2-5}
& \textbf{No reason}  & A label could not be assigned (due to lack of information). & \textbf{12} & \textbf{(6\%)}       \\
\midrule
\multirow{10}{*}{\rotatebox{90}{Purpose}} &
Document for later fix & Document an issue that should be revisited in the future. & 68 & (34\%)      \\
\cmidrule{2-5} 
& Warning for future developers    & Warn other developers to pay attention to an aspect of the solution that may not be clear from its structure or content.   & 52 & (26\%)\\
\cmidrule{2-5} 
& Document suboptimal implementation choice & Explain why a problematic solution has been adopted.  & 46 & (23\%)\\
\cmidrule{2-5} 
& Document workaround            & Explicitly document constraints imposed by design or implementation choices. The comment contains workaround-related keywords, such as ``workaround'' and ``temporary''.   & 18 & (9\%)\\
\cmidrule{2-5} 
& Placeholder for later extension & Document an extension point for later enhancement(s).          & 12 & (6\%)\\
\cmidrule{2-5}   
& Silence build warnings & Defer or ignore warnings emitted by underlying tools. & 4 & (2\%) \\ 
\bottomrule
\end{tabular}}
\end{table*}

\smallskip
\noindent
\textbf{\emph{\underline{Results.}}}
Table~\ref{tab:code} provides an overview of the definitions and frequencies for three dimensions. We identified the 11 locations that span three categories, 17 reasons that fit into eight categories, and six purposes from our qualitative analysis.

\ul{\textit{More than half of pervasively cloned SATD comments configure external factors.}}
The upper portion of Table~\ref{tab:code} quantifies the locations that we observe for pervasively cloned SATD comments.
We observe that 58\% of SATD comments occur in the \texttt{Externals}, which configure factors that are external to the projects. Upon closer inspection, \texttt{Platform configuration} is the main external factor, accounting for 23\% of SATD comments. Example \ref{example2} presents a case of this location where statements contain conditional operator with \texttt{case \$host\_os in} to check the targeting operating system.
\begin{lstlisting}[linewidth=\columnwidth,breaklines=true, showstringspaces=false, columns=fullflexible,basicstyle=\footnotesize, caption={The SATD comment occurs in Platform configuration.}, label={example2}]
AC_MSG_CHECKING([whether the $compiler linker ($LD) supports shared libraries])
    _LT_TAGVAR(ld_shlibs, $1)=yes
    case $host_os in
      aix3*)
        # FIXME: insert proper C++ library support
        _LT_TAGVAR(ld_shlibs, $1)=no
\end{lstlisting}

The second most frequently occurring location is the \texttt{Behavioural} (30\% of coded SATD comments), which configures the behavior of the build system, such as environmental variables and flags. \texttt{Dynamic settings} is the main subcategory in this location, where statements define flags or environmental variables in build systems.  
Example~\ref{example3} presents an instance of \texttt{Dynamic settings}, which indicates the configuration of an undefined flag, i.e., ``-berok''.
\begin{lstlisting}[linewidth=\columnwidth,breaklines=true, showstringspaces=false, columns=fullflexible,basicstyle=\footnotesize, caption={The SATD comment occurs in Dynamic settings.}, label={example3}]
if test aix,yes = "$with_aix_soname,$aix_use_runtimelinking"; then
	# Warning - without using the other runtime loading flags (-brtl),
	# -berok will link without error, but may produce a broken library.
    _LT_TAGVAR(allow_undefined_flag, $1)='-berok'
\end{lstlisting}

\ul{\textit{Limitations in tools and libraries, as well as issues with their configuration are most frequent reasons for pervasively cloned SATD comments.}}
The middle portion of Table~\ref{tab:code} shows the frequency of SATD reasons. Same to our prior work~\citep{9551792}, the \texttt{Limitation} is the main reason for SATD comments (i.e., 38\%). These limitations include constraints imposed by the design or implementation of third-party libraries
or development tools. Unlike most SATD comments in Maven build system~\citep{9551792} due to \texttt{External library limitation}, we find that these three subcategories are diversely shared by limitations from tools (16\%), libraries (11\%), and build tool itself (11\%). 


Moreover, we find that \texttt{Configuration} is the second most frequent reason for SATD comments, accounting for 37\% of SATD comments. Coping those SATD is to annotate configuration issues during the compilation and build process. Example~\ref{example4} shows that this SATD comment occurred because of configuring the compiler version for different architectures.

\begin{lstlisting}[linewidth=\columnwidth,breaklines=true, showstringspaces=false, columns=fullflexible,basicstyle=\footnotesize, caption={The SATD comment occurs due to Compiler configuration.}, label={example4}]
# TODO: Check compiler version to see the suffix should be <arch>/gcc4.1 or
    #       <arch>/gcc4.1. For now, assume that the compiler is more recent than
    #       gcc 4.4.x or later.
    if(CMAKE_SYSTEM_PROCESSOR STREQUAL "x86_64")
      set(TBB_LIB_PATH_SUFFIX "lib/intel64/gcc4.4")
    elseif(CMAKE_SYSTEM_PROCESSOR MATCHES "^i.86$")
      set(TBB_LIB_PATH_SUFFIX "lib/ia32/gcc4.4")
\end{lstlisting}

We also discover new SATD reasons for sanitizing projects or post-install components. Example~\ref{example5} describes an instance of hard-coding the default installation path when releasing the project.
\begin{lstlisting}[linewidth=\columnwidth,breaklines=true, showstringspaces=false, columns=fullflexible,basicstyle=\footnotesize, caption={The SATD comment occurs due to releasing the project.}, label={example5}]
IF( ${DARWIN} )
    # TODO: set to default install path when shipping out
    SET( ALEMBIC_ROOT NOTFOUND )
\end{lstlisting}


\ul{\textit{Developers often introduce pervasively cloned SATD comments for documenting issues to be fixed later.}}
The bottom portion of Table~\ref{tab:code} presents the frequency of SATD purposes. These purpose categories perfectly matched our studied dataset. We observe that 34\% of SATD comments are left by developers with the purpose of \texttt{Document for later fix}. Example \ref{example5} describes an instance of this purpose where documenting the issue of the install path (i.e., ``\# TODO: set to default install path when shipping out''). The second frequent purpose is \texttt{Warning for future developers}, accounting for 26\% of SATD comments. Developers copy SATD comments to warn other developers to pay attention to an aspect of the solution that may not be clear from its structure or content.

\begin{tcolorbox}
\textbf{RQ4 Summary:}
We identify three locations, eight reasons, and six purposes from 200 examples of the most pervasively cloned SATD comments in build systems. Pervasively cloned SATD comments often deal with external factors. Prevalent reasons include limitations in tools and libraries. Developers often copy these commonly cloned SATD comments with the purpose of documenting issues to be fixed later.
\end{tcolorbox}













\section{Discussion}
\label{dis}
We now discuss the implications and challenges, as well as threats to validity.

\subsection{Implications and Challenges}
\label{sec:imp}
\fix{
We outline the implications and challenges stemming from our study below and provide a detailed discussion in the subsequent paragraphs.
\\\textbf{Implications:}
\begin{itemize}
    \item Although cloning rates in Autotools tend to be lower than the other studied build systems\fix{~\citep{10.1145/2591062.2591181}}, at least when it comes to SATD, cloning appears to be a pervasive practice.
    \item Developers could clean the legacy SATD comments in their systems.
    \item Tool builders could spare extra effort on those TDs that are easily resolved in these most common places (e.g., platform and tool configuration) and reasons (e.g., configuration and dependency issues).
    \item The innate property of SATD comments is used as a short-term memo for the technical debt of constraints to revisit in the future.
\end{itemize}
\textbf{Challenges:}
\begin{itemize}
    \item In future research, we intend to explore the relationship between the lifespan of SATD comments and the pervasiveness of SATD clones.
    \item A more thorough analysis of the relationship between technical debt and the comments that annotate it would likely shed light on the study of technical debt in future work.
    \item Researchers could consider removing uninformative SATD in a more strict constraint for SATD clone analysis.
    \item SATD in Java build systems is more likely to contain issue reports, and exposes common issues in C-Family build systems for future research.
    \item In future work, we plan to investigate the different cloning behaviors (file clones and commit clones) and their implications. 
    \item Researchers could propose an automatic awareness mechanism to remind those who want to reuse statements that contain possible technical debt.
\end{itemize}
}

\fix{\textbf{What is the relationship
between the lifespan of SATD comments and the pervasiveness of SATD clone?} In \textbf{RQ1}, we observed at least 62\% of SATD comments are clones in build systems. We suspect that
the high prevalence of clones could be attributed to the lifespan of SATD comments, i.e., the longer an SATD remains in the build systems, the greater the likelihood of
its clone being disseminated and distributed externally. In future research, we intend
to \textit{explore the relationship between the lifespan of SATD comments and the pervasiveness of SATD clones}. 
} 

\textbf{Why are clones of SATD comments in Autotools the most pervasive?} \citet{10.1145/2591062.2591181} analyzed the build system clones of the same set of build tools, observing that the more recent technologies (Maven and CMake) tend to be more prone to cloning than the older ones (Ant and Autotools). In \textbf{RQ1}, we observe that the clone proportion of SATD comments in Autotools (95\%) is the largest among the studied build tools. Moreover, Table~\ref{tab:system} shows that 30,937 SATD comments are distributed in 1,048 groups, accounting for 29.41 SATD comments per group. As well as statement clones, we find that similarity scores of statements surrounding SATD comments in Autotools are higher than other build systems in \textbf{RQ2}. \textit{Although cloning rates in Autotools tend to be lower than the other studied build systems\fix{~\citep{10.1145/2591062.2591181}}, at least when it comes to SATD, cloning appears to be a pervasive practice.} 

\textbf{Why are SATD comments more prone to serve different statements?}
In \textbf{RQ2}, we find that the similarity scores of statements surrounding SATD clones are relatively lower than non-SATD clones. This result may indicate that SATD comments are more prone to serve different statements. As illustrated in the motivating example, this SATD propagated 5,413 occurrences in build systems. However, it may not necessarily describe the same hack. \textit{A more thorough analysis of the relationship between technical debt and the comments that annotate it would likely shed light on this in future work.}

\textbf{What are the reasons behind 45 groups of SATD clones that span multiple build tools?} 
\fix{
In \textbf{RQ1}, we conducted an additional analysis to delve deeper into the reasons for SATD clones across both build tools (CTC) and programming languages (CLC), as detailed in Table~\ref{tab:cross}. Our findings revealed that the majority of SATD clones spanning programming languages are attributed to annotation templates (accounting for 61\%) and annotation closures (comprising 34\%). \textit{Researchers could consider removing these uninformative SATD in a more strict constraint for SATD clone analysis}. Furthermore, we noticed that a significant portion of SATD clones across build tools point to prevalent issues in Autotools and CMake. Interestingly, some SATD clones refer to specific issue reports. This observation implies that \textit{SATD in Java build systems is more likely to contain issue reports}, and \textit{exposes common issues in C-Family build systems for future research}}

\textbf{Are these SATD clones due to file clones?} We find that large clone proportions of SATD
and non-SATD comments are cloned in \textbf{RQ1}. Moreover, in \textbf{RQ2}, we find that the similarity scores of statements surrounding SATD and non-SATD comments are high. In \textbf{RQ3}, we observe that the commits SHAs that introduce SATD clones could also be identical in different projects. \fix{
We suspect that developers may clone
entire blocks of build code without noticing the presence of
SATD comments from the results of extended analysis on similarity scores.} \textit{In future work, we plan to investigate these different cloning behaviors (file clones and commit clones) and their implications.}

\textbf{Legacy SATD comments in build systems.} In \textbf{RQ3}, \fix{our analysis on time interval (commits) of SATD clones could exist 1,548 commits on the median. However, technical debt is introduced for short-term goals. \citet{bavota2016large} found that SATD comments in source code linger for 266 commits on the median. These legacy SATD comments in build systems could create confusion for anyone inspecting the code. Thus, we recommend \textit{developers to clean these legacy SATD comments in their systems}. Finally, we recommend \textit{researchers to propose an automatic awareness mechanism to remind those who want to reuse statements that contain possible technical debt.}}

\textbf{Common locations, reasons, and purposes for pervasively cloned SATD.}
We characterize pervasively cloned SATD comments in \textbf{RQ4}, and we find that external factors (e.g., platform and tool configuration) are the most common location. We suspect that C-Family build systems need to define the target operating systems or other platform factors in their build systems, whereas Java build systems rely on the Java Virtual Machine (JVM) to abstract away the underlying external platform factors.

We observe that \texttt{Configuration} is the second most common reason in our study context. In our prior work on the Maven build system~\citep{9551792}, the \texttt{Dependency} reason was much more prominently featured.
We suspect that C-Family build systems are responsible for checking for the availability of libraries and features, whereas Java build systems have less build-time uncertainty for which they must account. For example, Maven doubles as a dependency management tool, which resolves missing dependencies in a uniform manner. Therefore, we suggest \textit{tool builders spare extra effort on those TDs that are easily resolved in these most common places (e.g., platform and tool configuration) and reasons (e.g., configuration and dependency issues).} 

Finally, we find that (i) limitations in tools and libraries are the most frequent reasons for pervasively cloned SATD, and (ii) developers often copy SATD comments with the purpose of documenting issues to be fixed later. These results are the same as SATD in Maven build system~\citep{9551792}, which indicates that \textit{the innate property of SATD comments is used as a short-term memo for the technical debt of constraints to revisit in the future.}



\subsection{Threats To Validity}
\label{threats}
\textbf{
Construct Validity.} 
\fix{Three} primary threats are summarized.
First, during the SATD comment extraction, we elected to apply a keyword-based approach as keywords are likely to generally summarize regularly occurring concepts. 
To ensure the reliability of the elected approach, we also evaluated the performance of the SATD detector~\citep{Liu}, a machine learning tool designed to identify SATD in source code. 
Similarly, we conducted a manual examination of 384 representative samples to verify comments that were not identified as SATD by the detector.
We observed that 27 comments (7\%) were missed by SATD detector, which is relatively higher than the keyword-based approach (5.4\%). 
Moreover, another limitation of the detector is that its training dataset is based on the annotated SATD comments in the context of source code instead of build specifications. \minor{Additionally, several state-of-the-art techniques have been proposed to identify SATD comments, such as the approach by \citet{guo2019mat}. However, we are confident that the keyword-based approach is reliable enough to be employed, as confirmed by the sanity check.}

Second, to gain a global picture of SATD comments in build systems, we select build systems that are adopted in Java and C/C++/Objective C projects. However, we only obtain 337 comments in the Ivy build system due to the popularity of Ivy. In the later SATD clones identification, none of these comments survived. Moreover, the dataset of SATD clones in each build system is unbalanced since we identified clones from different dimensions. Thus, a sampling of these SATD comments is not feasible. To mitigate this threat, we report our measurements in the density, percentage, or box plots (i.e., proportions of clones in \textbf{RQ1} and UAD or MCD metrics in \textbf{RQ3}).

\fix{Lastly, we intentionally prepared non-SATD comments from the pool of 2,565,906 non-SATD comments by obtaining two non-SATD comments that surround SATD comments. 
In contrast \fix{to the random sampling strategy, our method is likely to introduce a threat due to the locations of SATD and non-SATD comments. 
However, by using a random sampling method, we highly risk collecting a disproportionate number of non-SATD comments from frequently appearing build files, while neglecting comments from build files that are less common.
Thus, we chose to study the neighboring non-SATD comments. \minor{Moreover, the results may be biased by the selected number of non-SATD comments. 
To ensure proximity and similar program contexts shared with SATD ones, we chose to extract the non-SATD comments that are closest in position (one above and one below).
Nonetheless, as an exploratory study, our results do shed light on the cloning practices between SATD and non-SATD comments.} }}


\textbf{Internal Validity.} 
First, we apply \texttt{ANTLR4} lexical analyzers to extract comments from Autotools and CMake build systems. However, the official \texttt{ANTLR4} grammar for these build files is not found. With our customized \texttt{ANTLR4} grammars, it is possible to include false positives and miss uncovered comments in a build file that has syntax errors. However, these should be corner cases by our inspection.
Second, we did not manually code the whole dataset of SATD comments in build systems in \textbf{RQ4}, which will bring the risk of undiscovered SATD characteristic categories. To address this threat, we strive for theoretical saturation~\citep{eisenhardt1989building} that is widely adopted in
the SE domain~\citep{rigby2011understanding,zanaty2018empirical,hirao_fse, 9551792} to achieve analytical generalization. We performed three iterations and
achieved saturation after coding 200 SATD comments. \fix{Furthermore, we utilize \texttt{git log} combined with the \texttt{-L} option, specifically \texttt{git log ----pretty=format:\%H,\%an,\%ae,\%at,\%B ----no-patch -L start,end:filename}, to determine the authorship of SATD comments based on line numbers. Given that the \texttt{--follow} option is only applicable to an individual file and not a line range within that file, there is a potential we might overlook commits introducing SATD comments when ascertaining authorship for \textbf{RQ3}.}

\textbf{External Validity.} Although we obtained a large-scale and diverse dataset in four build systems from 6,502 projects, our results may not be generalized to other build systems (e.g., QMake and Ninja). Nonetheless, replication studies may help to improve the
strength of generalizations in future work.

\section{Related work}
\label{related}

\textbf{Self-Admitted Technical Debt.}
As an annotation of technical debt, the detection of SATD is also widely studied. \fix{\citet{maldonado2015detecting} extended the TD types from \citet{alves2014towards} to classify SATD in source code into design, defect, documentation, requirement, and test SATD. }
\citet{Liu} proposed the SATD detector to automatically detect and manage SATD comments in an Integrated Development Environment (IDE).
\citet{da2017using} automatically identified design and requirement SATD in source code comments by using NLP maximum entropy classifiers~\citep{manning2003optimization}. Moreover, \citet{ren2019neural} used Convolution Neural Network-based approaches with baseline text-mining approaches~\citep{keyword} to identify SATD in a cross-project prediction setting.
\citet{wait_for_it,9252045} identified ``On-Hold'' SATD (i.e., debt that a developer is waiting for a certain event or an updated functionality having been implemented elsewhere) for automated management.

\citet{Sierra} surveyed research work on SATD, analyzing the characteristics of current approaches and techniques for SATD detection, comprehension, and repayment. They suggest that researchers expand studies of SATD beyond the source code. The same suggestion is also proposed by the study on SATD
between industry and open-source systems developers~\citep{zampetti2021self}. Thus, researchers recently studied the presence of SATD in other software artifacts, i.e., build systems~\citep{9551792}, issue reports~\citep{xavier2020beyond, xavier2022documentation}, and code reviews~\citep{kashiwa2022empirical}. In our study, we use non-SATD as a baseline to better conclude the unique property of SATD clones. Although \citet{wehaibi2016examining} found that SATD do not have a clear relationship with defects, \citet{miyake2017replicated} replicated their study and agreed with the results of \citet{wehaibi2016examining}. Furthermore, SATD were found more effective to identify fix-prone
files and methods than non-SATD \citep{miyake2017replicated}.

 The closest work to ours is conducted by \citet{10.1145/3524610.3528387}, where they explored the existence and root causes of SATD clones in source code comments. 
Differently, our work systematically investigates the SATD clones in another core software development component namely build systems including the prevalence, the authorship, and their characteristics.

\textbf{Duplicate Software Artifacts.}
Researchers widely studied duplicate software artifacts for various purposes. Code clone is a common practice in software development. \citet{roy2007survey}, as well as \citet{koschke2007survey} performed surveys on the literature
on software clones. They discussed the definition and taxonomies of clones, clone detection techniques, and their evaluation. Other than clone detection, researchers observed bug-proneness~\citep{li2012cbcd} and bug propagation~\citep{mondal2019empirical} in code clones. Also, there are studies about cloning beyond source code~\citep{juergens2011research}. \citet{van2020clone} conducted an exploratory study concerning test code clones, and they found test code clones are more frequently occurring than source code. \citet{10.1145/2591062.2591181} observed similar results for build system clones in CMake/Autotools (C/C++) and Ant/Maven (Java). \citet{tsuru2021type} proposed a clone detection technique for deriving patterns within Dockerfiles. Software artifacts are not limited to these source code-like representations, modern software development has a broader scope than code, i.e., duplicate issue reports~\citep{bettenburg2008duplicate}, Q\&A posts~\citep{ahasanuzzaman2016mining}, and code reviews~\citep{hong2022commentfinder, wang2021automatic, li2020redundancy}. 

Our paper firstly focuses on cloning natural language text (i.e., SATD comments) in build systems, complementing the knowledge gap lying in the clone and SATD community. 


\section{Conclusion}
\label{conclus}
In this paper, we perform an empirical study on the cloning of 50,608 SATD comments in four major build systems. Specifically, we (i) explore the prevalence of SATD clones; (ii) investigate the similarity of statements; (iii) analyze SATD clones' authorship; and (iv) manually classify pervasively cloned SATD comments in terms of locations, reasons, and purposes.

We observe that: (i) SATD comments are pervasively cloned in build systems, accounting for 62--95\% of obtained SATD comments;
(ii) most statements are cloned with SATD comments, accounting for similarity scores of at least 0.8; 
(iii) a quarter of SATD clones are introduced by the author of the original SATD statements;
and (iv)
the most frequent reasons behind pervasively cloned SATD comments are limitations in tools and libraries, as well as issues with their configuration, and developers often introduced SATD comments with the purpose of documenting issues to be fixed later. 
This paper also foresees many promising avenues for future work, such as distinguishing different SATD clone behaviors and the automatic recommendation system of repaying SATD from resolved SATD clones.
\\\\
\noindent \textbf{Acknowledgements} This work was supported by JSPS Grant-in-Aid for JSPS Fellows JP23KJ1589, JSPS KAKENHI Grant Numbers JP20H05706, JP23K16864, and JST PRESTO Grant Number JPMJPR22P6.
\\\\
\noindent \textbf{Data Availability} To support the open science, we publish
a full replication package~\citep{replicate} online, including all the datasets, manually labeled data, supplementary materials (e.g., additional results), and scripts. This replication package is also available at \url{https://github.com/NAIST-SE/SATDClonesInBuildSystem}.
%
\section*{Declarations}
\textbf{Conflict of Interests} The authors declare that Hideaki Hata, is a member of the EMSE Editorial Board. All co-authors have seen and agree with the contents of the manuscript and there is no financial interest to report.

\bibliographystyle{spbasic}      
\bibliography{main}   

\vspace{2\baselineskip}
{\setlength\intextsep{0pt}
\begin{wrapfigure}{l}{25mm}
    \includegraphics[width=1in,height=1.25in,clip,keepaspectratio]{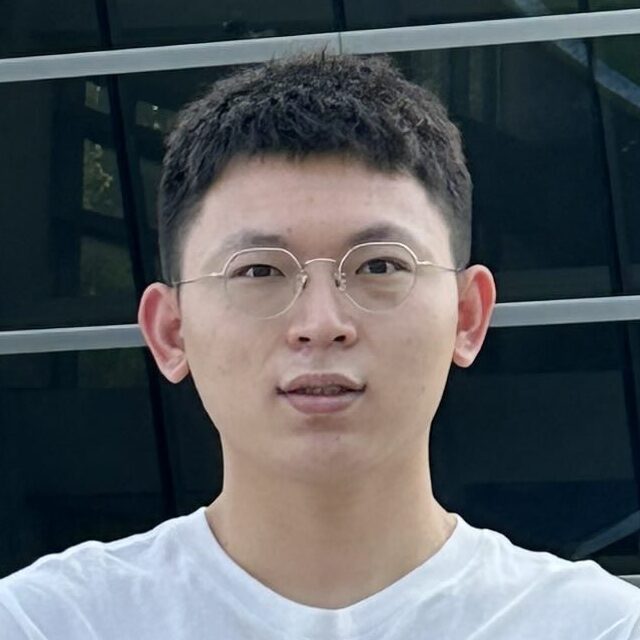}
\end{wrapfigure}\par
\noindent\textbf{Tao Xiao} is currently working toward a doctoral degree in the Department of Information Science, Nara Institute of Science and Technology, Japan. His main research interests include empirical software engineering, mining software repositories, and software testing. More about his work is available online at \url{https://tao-xiao.github.io/}. \par}
\vspace{2\baselineskip}

{\setlength\intextsep{0pt}
\begin{wrapfigure}{l}{25mm}
    \includegraphics[width=1in,height=1.25in,clip,keepaspectratio]{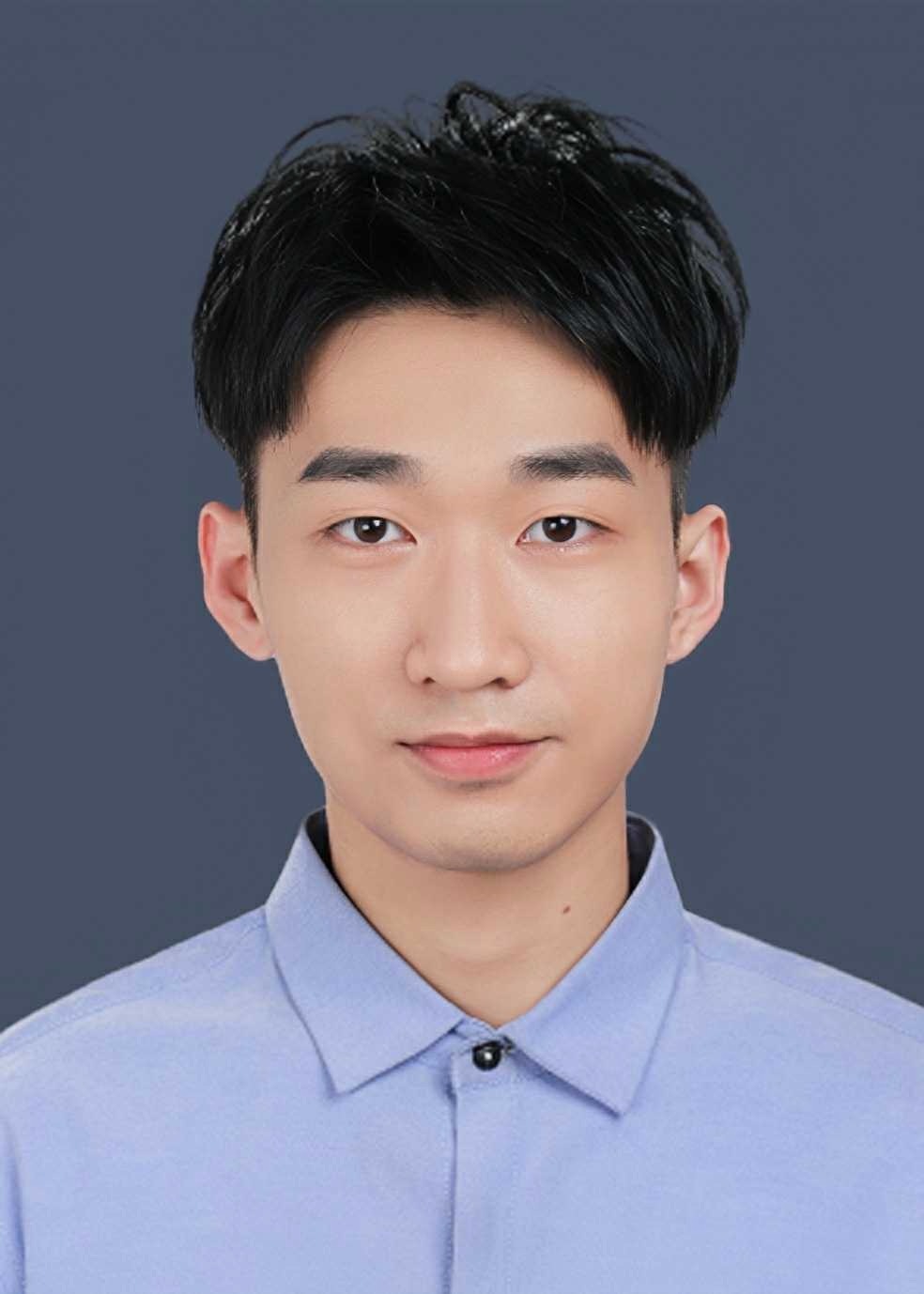}
\end{wrapfigure}\par
\noindent\textbf{Zhili Zeng}  is a master at the David R. Cheriton School of Computer Science, University of Waterloo, and a member of Software Repository Excavation and Build Engineering Labs (Software REBELs). His research interests lie in continuous integration (CI) acceleration and build systems. \par}
\vspace{2\baselineskip}

{\setlength\intextsep{0pt}
\begin{wrapfigure}{l}{25mm}
    \includegraphics[width=1in,height=1.25in,clip,keepaspectratio]{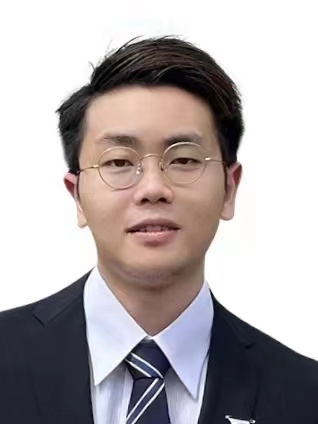}
\end{wrapfigure}\par
\noindent\textbf{Dong Wang} is a tenure-track Associate Professor at Tianjin University, China. 
Prior to this, he was an assistant professor at Kyushu University, Japan. He received Ph.D. degree from Nara Institute of Science and Technology, Japan. He is a member of the IEEE. His research interests include mining software repositories, empirical software engineering, and human aspects. More about his information is available online at \url{https://dong-w.github.io/}. \par}
\vspace{2\baselineskip}

{\setlength\intextsep{0pt}
\begin{wrapfigure}{l}{25mm}
    \includegraphics[width=1in,height=1.25in,clip,keepaspectratio]{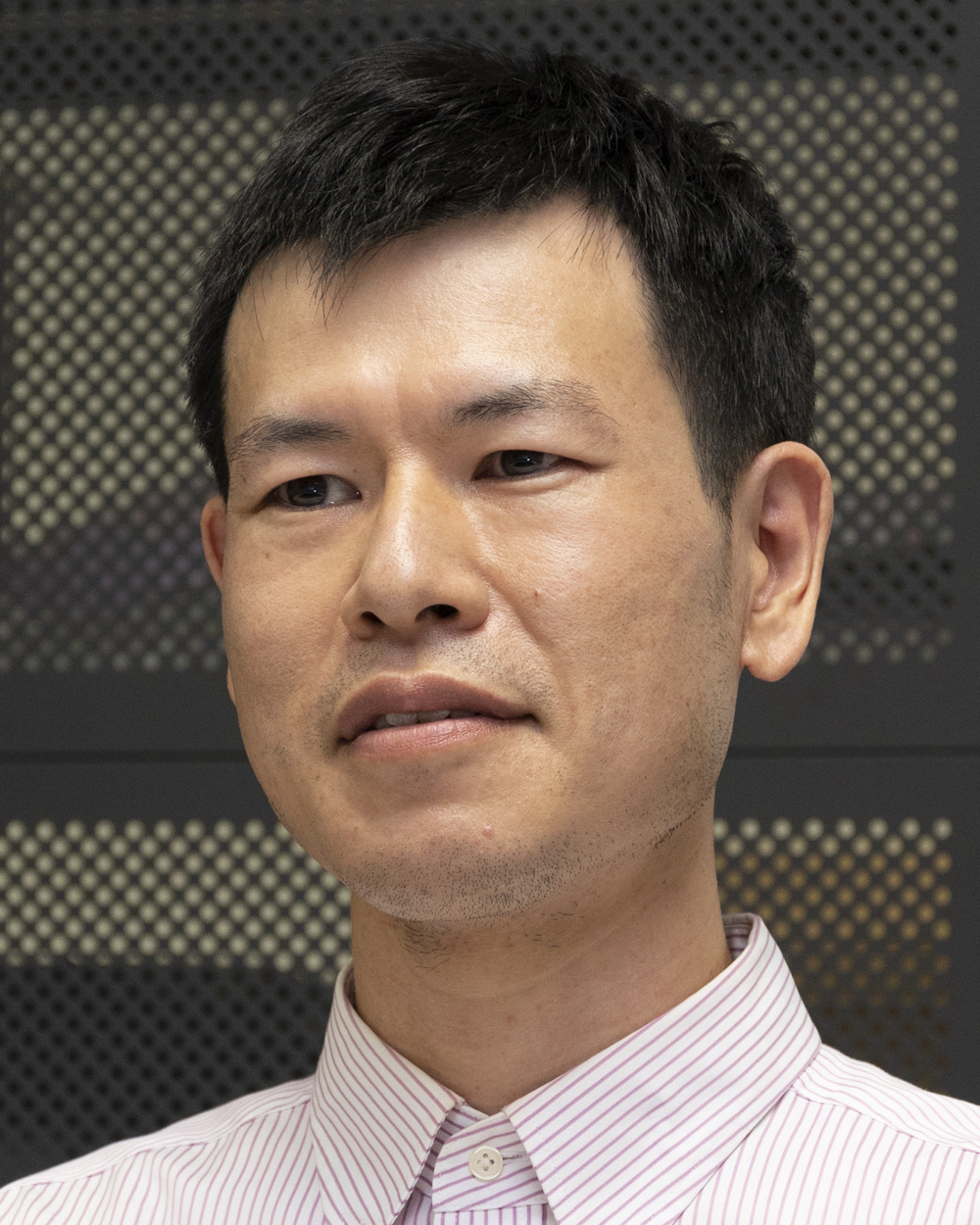}
\end{wrapfigure}\par
\noindent\textbf{Hideaki Hata} is an Associate Professor at Shinshu University. He received his Ph.D. in information science from Osaka University. His research interests include software ecosystems, human capital in software engineering, and software economics. More about Hideaki and his work is available online at \url{https://hideakihata.github.io/}. \par}
\vspace{2\baselineskip}

{\setlength\intextsep{0pt}
\begin{wrapfigure}{l}{25mm}
    \includegraphics[width=1in,height=1.25in,clip,keepaspectratio]{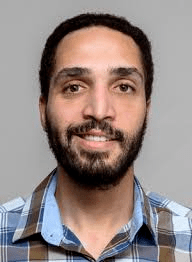}
\end{wrapfigure}\par
\noindent\textbf{Shane McIntosh} is an Associate Professor at
the University of Waterloo. Previously, he was an
Assistant Professor at McGill University, where
he held the Canada Research Chair in Software Release Engineering. He received his Ph.D. from Queen's University, for which he was awarded the Governor General's Academic Gold Medal. In his research, Shane uses empirical methods to study software build systems, release engineering, and software quality: \url{http://shanemcintosh.org/}. \par}
\vspace{2\baselineskip}

{\setlength\intextsep{0pt}
\begin{wrapfigure}{l}{25mm}
    \includegraphics[width=1in,height=1.25in,clip,keepaspectratio]{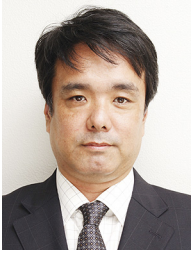}
\end{wrapfigure}\par
\noindent\textbf{Kenichi Matsumoto}
received the B.E., M.E., and Ph.D. degrees in Engineering from Osaka University, Japan, in 1985, 1987, 1990, respectively. Dr. Matsumoto is currently a Professor in the Graduate School of Information Science at Nara Institute Science and Technology, Japan. His research interests include software measurement and software process. He is a senior member of the IEEE and a member of the IPSJ and SPM. \par}
\vspace{2\baselineskip}
\end{document}